\documentclass[useAMS,usenatbib]{mn2e}
\usepackage{epsfig}
\usepackage{aas_macros}     
\usepackage{amsfonts}
\title[Simulations of Quintessential Cold Dark Matter: beyond the cosmological
constant]{Simulations of Quintessential Cold Dark Matter: beyond the cosmological
constant}
\author[E.~Jennings, C.~M.~Baugh, R.~E.~Angulo, S.~Pascoli]
{E. Jennings$^{1,2}$\thanks{E-mail: elise.jennings@durham.ac.uk}, C. M. Baugh$^{1}$, R. E. Angulo$^{3}$, S. Pascoli$^{2}$\\
$^{1}$ Institute for Computational Cosmology, Department of Physics, University of Durham, South Road, Durham, DH1 3LE, U.K.\\
$^{2}$ Institute for Particle Physics Phenomenology, Department of Physics, University of Durham, South Road, Durham, DH1 3LE, U.K.\\
$^{3}$  Max Planck Intitute fur Astrophysik, D-85741 Garching, Germany.\\
}
\begin{document}

\date{}

\pagerange{\pageref{firstpage}--\pageref{lastpage}} \pubyear{2009}

\maketitle

\label{firstpage}

\begin{abstract}
We study the nonlinear growth of cosmic structure in different dark energy
 models, using large volume N-body simulations.
 We consider a range of quintessence models
 which feature both rapidly and slowly varying dark energy equations of state, 
and compare the growth of structure to that in a universe
 with a cosmological constant.
 We use a four parameter equation of state 
for the dark energy which accurately reproduces the 
quintessence dynamics over a wide range of redshifts.
The adoption of a quintessence model changes the expansion history of the universe, the form of 
the linear theory power spectrum and can alter key observables, such as the horizon scale and the 
distance to last scattering. We incorporate these effects into our simulations in stages to isolate the impact of each on the 
growth of structure.
The difference in structure formation can be explained to first order by the difference in growth factor at a given epoch; 
this scaling also accounts for the nonlinear growth at the 15\% level.
We find that quintessence models
that are different from $\Lambda$CDM both today and at high redshifts $(z \sim 1000)$ and which feature 
late $(z<2)$, rapid transitions in the equation of state,
  can have identical baryonic acoustic oscillation (BAO) peak positions
to those in $\Lambda$CDM.
We find that these models  have higher 
abundances of dark matter haloes at $z>0$ compared to $\Lambda$CDM and so measurements of the mass function should allow us to distinguish 
these quintessence models from 
a cosmological constant. 
However, we find that a second class of quintessence models, whose equation of state makes an early $(z>2)$ rapid transition to $w=-1$, cannot be distinguished 
from $\Lambda$CDM using measurements of the mass function or the BAO, even if these models have non-negligible amounts of dark energy at early times.

\end{abstract}

\begin{keywords}
Methods: N-body simulations - Cosmology: theory - large-scale structure of the Universe
\end{keywords}

\section{Introduction}

Determining whether or not the dark energy responsible for the accelerating expansion of the Universe evolves with time remains a key goal of physical cosmology.
 This will tell us if the dark energy is indeed a cosmological constant or has a 
dynamical form as in quintessence models.
The nature of the dark energy determines the expansion history of the Universe and hence the rate at which cosmological perturbations grow. In this paper we investigate the 
influence of quintessence dark energy on the nonlinear stages of structure formation using a suite of N-body simulations.

The simplest candidate for dark energy is the cosmological constant, $\Lambda$ 
(see e.g. the review by \citealt{Carroll:2000fy}). 
Despite the success of $\Lambda$CDM (cold dark matter and cosmological constant model) at fitting much of the available observational data 
\citep{Sanchez:2009jq},  this model fails to address 
two important issues, the fine tuning problem and the coincidence problem. 
The fine-tuning problem arises from the 
 vast discrepancy between the
vacuum energy level predicted by particle physics,
generically given by $\Lambda^4$, where $\Lambda$ is the physics scale considered, and the  value of missing energy density
inferred cosmologically, $\rho \sim 10^{-47} \rmn{GeV}^4$. In the standard model of particle physics, $\Lambda$ could be at the Planck scale, $\Lambda \sim 10^{18}$GeV.
The coincidence problem refers to the fact that
 we happen to live around the time at which dark energy has emerged as the dominant component of the Universe, and has a comparable energy density 
to matter, $\rho_{\tiny \mbox{DE}} \sim \rho_m$. 

Quintessence models were devised to solve the fine tuning and coincidence problems of $\Lambda$CDM.
In these models, the cosmological constant  is replaced by an extremely light scalar field
which evolves slowly \citep{Ratra:1987rm,1988NuPhB.302..668W,1998PhRvL..80.1582C,Ferreira:1997hj}. 
An abundance of quintessence models has been proposed in the literature which 
can resolve the coincidence problem and explain the observationally inferred amount of dark energy.
Models of quintessence dark energy can have very different potentials, $V(\varphi)$, but can share common features.  The potentials provide the correct magnitude 
of the energy density and are able to drive the accelerated expansion seen today. The form of the 
scalar field potential  determines the trajectory of the equation of state, $w(z)=P/\rho$, as it evolves in time.
Hence, different quintessence dark energy models have different dark energy densities as a function of time, $\Omega_{\tiny \mbox{DE}}(z)$.
This implies a different growth history for dark matter perturbations from that expected in $\Lambda$CDM. 

Cosmological N-body simulations are the theorist's tool of choice for modelling the final stages of perturbation collapse.
The overwhelming majority of simulations have used the concordance $\Lambda$CDM
cosmology. Here we simulate different dark energy models and study their 
observational signatures.
A small number of papers have used N-body simulations to test 
scalar field cosmologies 
 \citep{ Ma:1999dwa,Linder:2003dr,Klypin:2003ug, Francis:2008md,Grossi:2008xh,2009JCAP...03..014C,2009arXiv0903.5490A}. 
Rather than explicitly solving for different potentials, it is standard practice to modify the Friedmann equation using a form for the dark energy
equation of state, $w(z)$.
Previous work 
used a variety of 
parametrizations for  $w(z)$, the most common being the two parameter  equation, $w = w_0 + (1-a)w_a$
 \citep{Chevallier:2000qy, Linder:2002et} or the empirical three parameter equation proposed by \citet{Wetterich:2004pv} for the so-called early dark energy models. 
The disadvantage of using a 1 or 2 variable parametrization for $w$ is that it cannot accurately reproduce the dynamics of 
a quintessence model over a wide range of redshifts.
If we wish  to 
reproduce the equation of state of the original scalar field to within 5\%,
a two parameter equation of state will not be able to achieve this precision for a wide range of quintessence potentials \citep{Bassett:2004wz}.  
 Instead, we take advantage
of a parametrization for $w(z)$ which can describe a wide range of different models.
In this work we use a four parameter dark energy equation of state  which can accurately reproduce the original $w(z)$
for a variety of dark energy models to better than 5\% for redshifts $z<10^3$ \citep{Corasaniti:2002vg}.

In this paper we present three stages of N-body simulations of structure formation in quintessence models. Each stage progressively relaxes the 
assumptions made and brings us closer to a full physical model.
 In the first stage, 
the initial conditions for each quintessence cosmology are generated using a $\Lambda$CDM linear theory power spectrum and 
the background cosmological parameters are the best fit values assuming a $\Lambda$CDM cosmology.
  The only departure from $\Lambda$CDM in this first stage is the dark energy equation of state and its impact on the expansion rate. In the second stage, 
 we use a modified version of CAMB \citep{Lewis:2002ah} to generate a consistent linear theory power spectrum for each quintessence model. The linear 
theory power spectrum  can differ from the power spectrum in $\Lambda$CDM due to the presence of non-negligible 
amounts of dark energy during the early stages of the matter dominated era. This power spectrum 
is then used to generate the initial conditions for the N-body simulation which is run again for each dark energy model. 
 The third and final stage in our analysis 
is to find the values for the  cosmological parameters, $\Omega_{\rm m} h^2$, $\Omega_{\rm b} h^2$ and $H_0$ (the matter density, baryon density and Hubble parameter)
such that each model satisfies
 cosmological distance constraints.
Recently  \citet{2009arXiv0903.5490A} used CMB and SN data to constrain the parameters in the quintessence potential and the value of 
the matter density, $\Omega_{\rm m} h^2$, for two models.
In this paper we allow three parameters to vary when fitting each quintessence model to the available data. 
This distinction is important as changes in these parameters may produce compensating effects which result in the quintessence model 
looking like $\Lambda$CDM.  For example, for a given dark energy equation of state, 
a lower value of the matter density may not result in large changes in the Hubble parameter if the value of $H_0$ is increased.
 In going through each of these stages we build 
up a comprehensive picture of the quintessence models and their effect on the nonlinear growth of structure.

This paper is organised as follows. In Section 2 we discuss quintessence models and the parametrization we use for the dark energy equation of state. 
 We also outline the expected impact of different dark energy models on structure formation. In Section 3 we give the  details 
of our N-body simulations. 
The main power spectrum results are presented in Section \ref{wmap}. Intermediate results are presented 
in Sections \ref{psI} and  \ref{psII}, which the reader may wish to omit on a first pass.
In Section \ref{mf}  we present the mass function predictions. 
In Section \ref{bao} we discuss the appearance of the baryonic acoustic oscillations in the matter power spectrum. Finally, in Section 6 we present our conclusions.

\section[]{Quintessence Models of Dark Energy \label{QUIN}}

Here we briefly review some general features of quintessence models; more detailed descriptions can be found, for example,  in 
 \citet{Ratra:1987rm,1988NuPhB.302..668W,Ferreira:1997hj,Copeland:2006wr} and \cite{Linder:2007wa}.
The main components of quintessence models are~radiation, 
pressureless matter and a quintessence scalar field, denoted by $\varphi$.
This dynamical scalar field is a slowly evolving component with negative pressure. 
This multifluid system can be described by the following action
\begin{eqnarray}
\rmn{S} = \int {\rm d}^4x \sqrt{-g}\,(-\frac{\rmn{R}}{2\kappa} + \mathcal{L}_{\tiny{\rmn{m +r}}}
+\frac{1}{2} \,g^{\mu \nu}\,\partial_{\mu}\varphi\,
\partial_{\nu}\varphi -V(\varphi)  ) \, ,
\end{eqnarray}
where $\rmn{R}$ is the Ricci scalar, $\mathcal{L}_{\rmn{m+r}}$ is the 
Lagrangian density of matter and radiation, $\kappa = 8\pi G$, $g$ is the 
determinant of  a spatially flat
Friedmann-Lema\^{i}tre-Robertson-Walker (FLRW) metric tensor $g_{\mu \nu}$ and  $V(\varphi)$ is the scalar field potential.
We assume that any couplings to other fields are negligible so that the scalar field interacts with other matter only through gravity.
Minimising the action with respect to the scalar field  
leads to its equation of motion 
\begin{equation}
\label{eqnofm}
\ddot{\varphi} + 3\,H\,\dot{\varphi} + \frac{{\rm d} V(\varphi)}{{\rm d}\varphi} = 0 \,,
\end{equation}
 where $H $ is the Hubble parameter and we have assumed the field is spatially homogeneous, $\varphi(\vec{x},t)=\varphi(t)$. 
The impact of the background on the dynamics of $\varphi$ 
is contained in the $3H\dot{\varphi}$ term.  
The Hubble parameter for dynamical dark energy 
in a flat universe is given by
\begin{equation} 
\frac{H^2(z)}{H_0^2}=\left( \Omega_{\rm m} \,(1+z)^{3} + (1-\Omega_{\rm m}) e^{3\int_0^z 
d {\tiny \mbox{ln}} (1+z') \, [1+w(z')]}\right),
\end{equation}
where $H_0$ and $\Omega_{\rm m} = \rho_{\rm m}/\rho_{{\tiny \mbox{crit}}}$ are the values of the  Hubble parameter 
and dimensionless matter density, respectively, at redshift $z =0 $ and $\rho_{{\tiny \mbox{crit}}} = 3H_0^2/(8\pi G)$ is the critical density.
The dark energy equation of state is expressed as the ratio 
of the dark energy 
pressure to its energy density, denoted as $w = \rmn{P}/\rho$.
Once a standard kinetic term is assumed in the quintessence model,  
it is the choice of potential
 which determines 
$w$ as
\begin{equation}
\label{wgeneral}
w = \frac{\dot{\varphi}^2/2 - V(\varphi)}{\dot{\varphi}^2/2 + V(\varphi)}\,.
\end{equation}
In general in these theories if the contribution from the kinetic 
($\dot{\varphi}=0$) and gradient energy (${\rm d}\varphi/{\rm d}\vec{x} =0$) is negligible,
then the effect of the scalar field  is equivalent to a cosmological constant which behaves as a perfect fluid, with
$ P = -\rho$ or $w=-1$.

\subsection{Classes of quintessence models}

Two broad classes of quintessence models can be used to solve both the fine-tuning and coincidence problems. 
The first is based on the idea of so called \lq tracker fields\rq \,\citep{Steinhardt:1999nw}. These fields adapt their behaviour 
to the evolution of the scale factor and hence track the background density. 
The other class is
 referred to as \lq scaling solutions\rq \, \citep{Halliwell:1986ja,1993NYASA.688..647W,Wetterich:1994bg}.
In these models the ratio of energy densities, $\rho_{\varphi}/\rho_{\rm B}$, is constant.

In tracking models, the $\varphi$ field rolls down its potential, $V(\varphi)$,
to an attractor-like solution.
The great advantage of these models is that this solution 
is insensitive to the initial conditions 
of the scalar field produced after inflation. 
A general feature of these  tracking solutions is that as the scalar field is tracking behind the dominant matter component in the universe, its
equation of state, $w_{\varphi}$, 
depends on the background component  as
\begin{equation}
\frac{\rho_{\varphi}}{\rho_{\rm B}} = a^{3\,(w_{\rm B}-w_{\varphi})}\, ,
\end{equation}
where $\rho_{\rm B}$ and $w_{\rm B}$ denote the background energy density and equation of state respectively,
with $w_{\rm B} = 1/3$ (radiation era) and $w_{\rm B} = 0$ (matter era).
As a result, the energy density of the scalar field remains sub-dominant during the
 radiation and matter dominated epochs, although it decreases at a slower rate than
the background density. The quintessence field,   $\rho_{\varphi}$, naturally
emerges as the dominant component today and its equation of state is driven towards $w=-1$.
An example of  a tracking model is the inverse potential form proposed by \citet{Zlatev:1998tr},
 $V(\varphi)\sim M^{4+\alpha}\varphi^{-\alpha}$,  where
 $M$ is a free parameter that is generally fixed by the requirement
that the dark energy density today $\Omega_{\tiny \mbox{DE}} \sim 0.7$ and so the quintessence potential must be
 $V \sim \rho_{{\tiny \mbox{crit}}}$. This
implies that $\varphi $ is of the order
 of the Planck mass today, $\varphi \sim M_{\rm Pl}$.
With  $\alpha \leq 6$, the quintessence field equation of state is approximately $w_0 \leq -0.4$ today.

In scaling quintessence models, the ratio of energy densities, $\rho_{\varphi}/\rho_{\rm B}$, is kept constant,  
unlike tracking models, where $\rho_{\varphi}$ changes more slowly than
 $\rho_{\rm B}$. 
During the evolution  of the energy density in a \lq scaling\rq \, model,
if the dominant matter component advances
as $\rho \propto a^{-n}$, then the scalar field will obey
$\Omega_{\varphi} = n^2/\alpha^2$
after some initial transient behaviour.
Scaling quintessence models can
suffer from an inability to produce
late time acceleration, whilst at the same time adhering to
observational constraints, such as, 
for example, the lower limit on $\Omega_{\varphi}$ during
nucleosynthesis \citep{Bean:2001wt}.
\citet{Albrecht:1999rm} used a modified coefficient in their scaling potential,
$V(\varphi) = V_{\rm p} \,e^{-\lambda\,\varphi}$, where $V_{\rm p}(\varphi) = (\varphi - B)^{\alpha} +A$,  resulting in
 a model which can produce late time acceleration as well as satisfying
cosmological bounds, for a variety of constants $A$ and $B$.
\citet{2000PhRvD..61l7301B} considered  a linear combination of exponential terms in the scalar field potential and found this yielded a larger range of acceptable initial
energy densities for $\varphi$ compared with inverse models.
\citet{Copeland:2000vh} also consider supergravity (SUGRA) corrections to quintessence models, where the resulting potential can exhibit either
 \lq tracking\rq \, or  \lq scaling\rq \, behaviour depending on which path the scalar field takes down its potential
towards the minimum where it would appear as a cosmological constant.

The physical origin of the 
quintessence field should be addressed by models motivated by 
high energy particle physics.
As the vacuum expectation value of the scalar 
field today is of the order of the Planck mass, any candidates for 
quintessence which arise in supersymmetric (SUSY) gauge theories 
may receive supergravity corrections which will alter the field's potential.
It is this fact that motivates many authors to argue that any quintessence model inspired by particle physics potentials must be based on 
SUGRA. 
\citet{Brax:1999gp} discuss such models and employ the potential $V(\varphi) = \Lambda^{4+\alpha}/\varphi^{\alpha}e^{\kappa/2\varphi^2}$ with 
a value of $\alpha \geq 11$
in order to drive $w_0$ close to $-1$ today.

\begin{figure}
{\epsfxsize=9.truecm
\epsfbox[101 359  469 718]{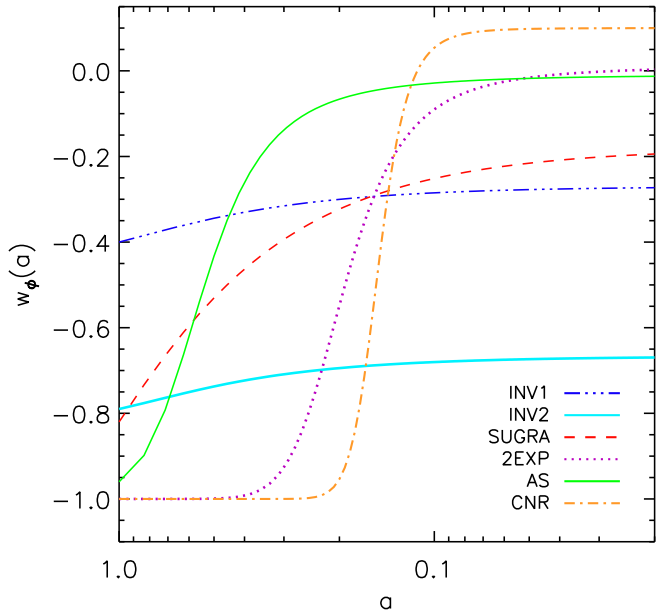}}
\caption{
The dark energy equation of state as a function of expansion factor, 
$w(a)$, for six quintessence models motivated by particle physics, which are either tracking or scaling solutions.
The parametrization for $w(a)$ is given in Eq. \ref{weq} and the four parameter values which specify each model are  given in Table \ref{wparam}. 
Note the left hand side of the x-axis is the present day.} \label{w}
\end{figure}

In summary, in this paper we will consider  six quintessence models which cover the 
behaviours discussed above.  In particular,
INV1 and INV2, 
which are  plotted in Fig. \ref{w},  have inverse power 
law potentials and exhibit tracking solutions. The INV1 model is the \lq INV\rq \, model
 considered by \citet{Corasaniti:2002vg} and has a value of $w_0 = -0.4$ today. 
As current observational data favour a value of $w_0 < -0.8$ \citep{Sanchez:2009jq}, the INV1 model will be used as an illustrative model.  
We shall consider a second inverse power law model (INV2) which is in better agreement with the constraints on $w$.
As noted by \citet{Corasaniti:2004rm},
 the scale $\Lambda$ in the inverse power law potential, $V(\varphi) = \Lambda^{\alpha +4}/\varphi^{\alpha}$
is fixed by the  value of $\Omega_{\mbox{\tiny{DE}}}$ today. Solving the coincidence problem  requires this scale for $\Lambda$  to be
consistent with particle physics models. For values of $\alpha \geq 6$ it is possible to have energy scales of  $\Lambda \sim 10^6$ GeV.
Setting $\alpha = 6$ results in an equation of state with $w_0 = -0.4$ 
(INV1). It is possible to drive the equation of state closer to $-1$
today with lower values of $\alpha$, although the value of $\Lambda$ is then pushed to
an undesirable energy range when compared with the typical scales of  particle
physics. The second model INV2,  which has  $w_0 = -0.79$ with $\alpha = 1$,
has been added to illustrate a power law potential with a dark energy equation of state which agrees
with constraints found on $w_0$ using CMB, SN and large scale structure data \citep{Sanchez:2009jq}.
We also use the SUGRA model of \citet{Brax:1999gp} which exhibits tracking field behaviour. The potential in this case also contains an
exponential term which pushes the  dark energy equation of state to $w_0 =-0.82$. 
The 2EXP model is an example of a scaling solution and features a double exponential term in the scalar field potential 
\citep{2000PhRvD..61l7301B}. 
The AS model suggested by \citet{Albrecht:1999rm} 
belongs to the class of scaling quintessence fields. As mentioned previously, the parameters in this potential can be adjusted 
to have the fractional dark energy density, $\Omega_{\tiny \mbox{DE}}$,  below the nucleosynthesis bound in the early universe. 
The CNR model  \citep{Copeland:2000vh} has a tracking potential where  the scalar field rolls down to its minimum and will settle down to
$w_0 = -1$ after a series of small oscillations.

 Each of the quintessence models we consider is one of a 
family of such models with parameter values 
chosen in order to solve the issues of fine-tuning and 
coincidence, as well as to produce a value of $w_0 \sim -1$ today. These requirements limit
 the parameter space available to a particular quintessence potential. 
 For example, this limits the range of the \citet{Brax:1999gp} SUGRA model. The SUGRA model we simulate has a fixed parameter value in the supergravity 
potential but the dark energy equation of state for this model does not depend strongly on this parameter (see Figure 4 in \citealt{Brax:1999gp}).

\subsection{Parametrization of $w$}

Given the wide range of quintessence models in the literature
 it would be a great advantage, 
when testing these models, to obtain one model independent equation describing the evolution of  the dark energy equation of state
without having to specify the potential $V(\varphi)$ directly. 
Throughout this paper 
we will employ the  parametrization for $w$ proposed by \citet{Corasaniti:2002vg}, which is a generalisation of the method used by \citet{Bassett:2002qu} for fitting dark energy models 
with rapid late time transitions.
Using a parametrization for the dark energy equation of state provides us with a model independent probe of several dark energy properties.
The dark energy equation of state, $w(a)$, is described by its value during radiation 
domination, $w_{\rmn{r}}$,
 followed by a transition to a plateau 
in the  matter dominated era, $w_{\rm m}$,
 before making the transition to the present
day value $w_0$. Each of these transitions can be parametrized by the 
scale factor $a_{\rmn{r},{\rm m}}$ at which they occur and 
the width of the transition $\Delta_{\rmn{r},{\rm m}}$. 

In order to reduce this parameter space we use the 
shorter version of this parametrization for $w$, 
which is relevant as our simulations begin in the matter dominated era. 
The equation for $w$ valid after matter-radiation equality is~  
\begin{equation}
\label{weq}
w_{\varphi} (a) = w_0 + (w_{\rm m} - w_0) \,\times \, \frac{1+e^{\frac{a_{\rm m}}{\Delta_{\rm m}}}}{1+e^{-\frac{a-a_{\rm m}}{\Delta_{\rm m}}}}\,\times \, \frac{1-e^{-\frac{a-1}{\Delta_{\rm m}}}}{1-e^{\frac{1}{\Delta_{\rm m}}}}\,.
\end{equation}
\citet{Corasaniti:2002vg} showed that this four parameter fit gives an excellent match to the exact equation of state.
Table \ref{wparam} gives the best fit values for the equation of state parameters
for the different quintessence models
taken from \citet{Corasaniti:2002vg}, with the addition of the  INV2  model.
The parametrization for the dark energy equation of state  is plotted in Fig. \ref{w} for the various quintessence models used in this paper.

\begin{figure}
{\epsfxsize=9.truecm
\epsfbox[101 359 469 718]{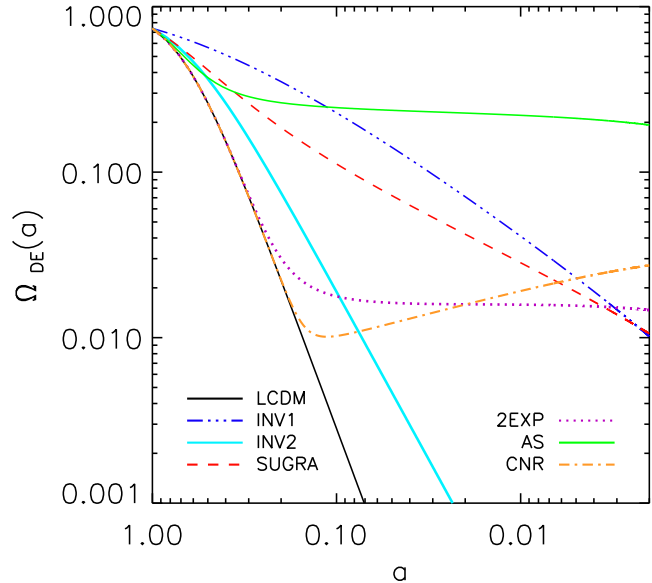}}
 \caption{The dark energy density, $\Omega_{\tiny \mbox{DE}}(a)$, as a function of expansion factor. The INV1, SUGRA, CNR,  2EXP and AS  
models have significant levels of dark energy at early times. 
From $z \sim 9$ until today the  2EXP and CNR 
models display the same energy density as $\Lambda$CDM. Note the x-axis scale on this plot goes to $z >300$ on the right hand side.}
\label{omegade} 
\end{figure}

\begin{figure}
{\epsfxsize=8.truecm
\epsfbox[101 359 469 718]{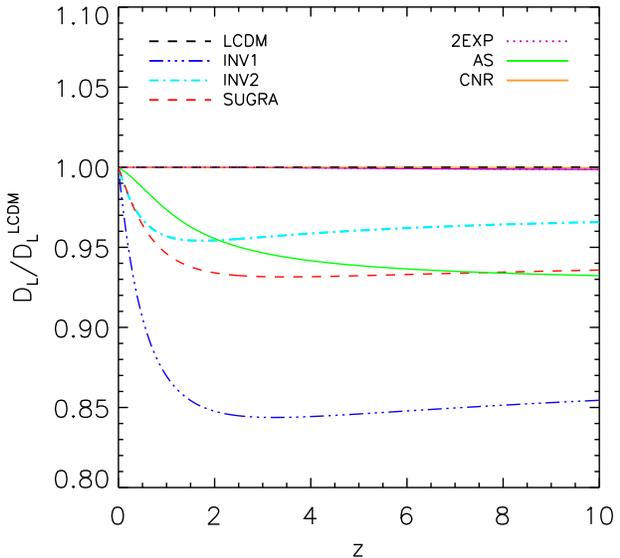}}
\caption{ The luminosity distance in different quintessence models 
compared to that in a $\Lambda$CDM cosmology. In this case we have assumed the same matter density today of 
$\Omega_{\rm m} =0.26$
in each of the models. The CNR and 2EXP models predict the same $D_L$ as in $\Lambda$CDM and are overplotted.} 
\label{dl}
\end{figure}

\begin{figure}
{\epsfxsize=8.truecm
\epsfbox[101 359 469 718]{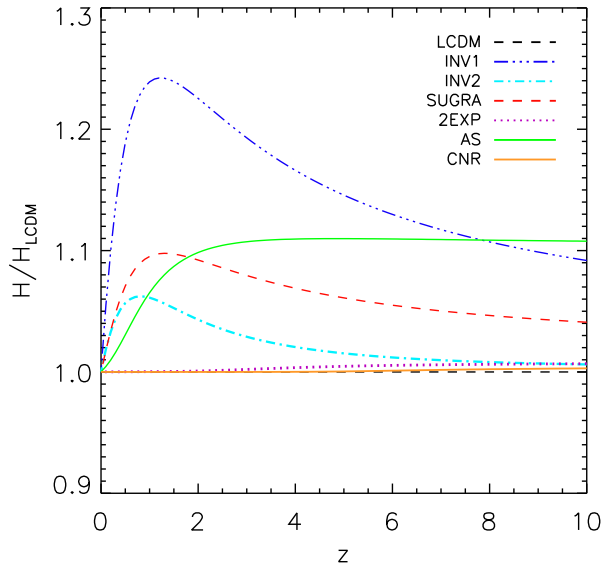}}
\caption{ The ratio of the Hubble parameter for quintessence cosmologies to that in $\Lambda$CDM. 
}
\label{Hubble}
\end{figure}

Fig. \ref{omegade} shows the evolution of the dark energy density with expansion factor in each quintessence model. 
Some of these models display significant levels of dark energy 
at high redshifts in contrast to a $\Lambda$CDM cosmology. As the AS, CNR, 2EXP and SUGRA models have non-negligible 
dark energy at early times, all of these could be classed as \lq early dark energy\rq \, models.
As shown in Fig. \ref{omegade} both the CNR and the 2EXP models have high levels of dark energy at high redshifts compared to $\Lambda$CDM; 
 after an early rapid transition, the dark energy density evolves in the same way as in a $\Lambda$CDM cosmology.
Other models, like the AS, INV1 and the SUGRA models, 
also have non-negligible amounts of dark energy at early times,
and after a late-time transition, the dark energy density mimics a $\Lambda$CDM cosmology at very low redshifts.
In Section 4 we will investigate if  quintessence models which feature an early or late transition in their equation of state, and in their dark 
energy density, can be distinguished from $\Lambda$CDM by examining  the growth of large scale structure.
 The luminosity distance and Hubble parameter in the quintessence models are
compared to $\Lambda$CDM in Fig. \ref{dl} and Fig. \ref{Hubble}, respectively.
In these plots it is clear that the CNR and the 2EXP models differ from $\Lambda$CDM only at very high redshifts.

The adoption of a 4 variable parametrization is essential to accurately model the expansion history over the full range of redshifts probed by the simulations.
Using a 1 or 2 parameter equation of state whose application is limited to low redshift measurements 
restricts the analysis  of the properties of dark energy and cannot make use of high redshift measurements such as the CMB. 
As an example,  \citet{Corasaniti:2004rm}  demonstrated that a two parameter log expansion
for $w(z)$ proposed by \citet{Gerke:2002sx}, can only take into account a quintessence model which varies slowly and cannot faithfully reproduce the original $w(z)$ at high redshifts. 
\citet{Bassett:2004wz} analysed how accurately various parametrizations 
could reproduce the dynamics of quintessence models. They found that parametrizations based on an 
expansion to first order in $z$ or $\mbox{log} \,z$ showed errors of $\sim 10\%$ at $z=1$.
A general prescription for $w(z)$ containing more parameters than a simple 1 or 2 variable equation can accurately describe both 
slowly and rapidly varying equations of state \citep{Bassett:2004wz}. For example, the parametrization provided by \citet{Corasaniti:2002vg} can accurately mimic the exact time behaviour of $w(z)$
to $<5\%$ for $z<10^3$  using a 4 parameter equation of state and to $<9\%$ for $z<10^5$ with a 6 parameter equation.
Finally, we note that the  parametrization for $w$ proposed by \citet{Corasaniti:2002vg}
is similar to the four parameter equation of state in \citet{Linder:2005ne} (Model 4.0) 
 where the evolution of $w$ is described in
terms of the e-fold variable, $ N = \mbox{ln}\,a$, where $a$ is the scale factor.

\setcounter{table}{0}
\begin{table}
\caption{The equation of state of the dark energy models simulated, expressed in the parametrization of \citet{Corasaniti:2002vg}.
The evolution of $w(a)$ is described by four parameters, the value of the equation of state today, $w_0$, and during matter domination era, $w_{\rm m}$, 
the expansion factor, $a_{\rm m}$, when the field changes its value during matter domination and the width of the transition, 
$\Delta_{\rm m}$.  We have added the  INV2 model to this list as an example of an inverse power law potential with a value of $w_0$ closer to -1 than in 
the INV1 model. }
\label{wparam}
\center
\begin{tabular}{|c|r|r|r|r|}
\hline \hline
Model & $w_0$ & $w_{\rm m}$ & $a_{\rm m}$ & $\Delta_{\rm m}$   \\
\hline
INV1 & -0.4 & -0.27 & 0.18 & 0.5 \\
INV2 & -0.79 & -0.67 & 0.29 & 0.4 \\
SUGRA &  -0.82 & -0.18 & 0.1 & 0.7  \\
2EXP &  -1.0 & 0.01 & 0.19 & 0.043  \\
AS & -0.96 & -0.01 & 0.53 & 0.13  \\
CNR & -1.0 & 0.1 & 0.15 & 0.016 \\
\hline
\end{tabular}
\end{table}

\begin{figure}
{\epsfxsize=8.truecm
\epsfbox[82 369 269 702]{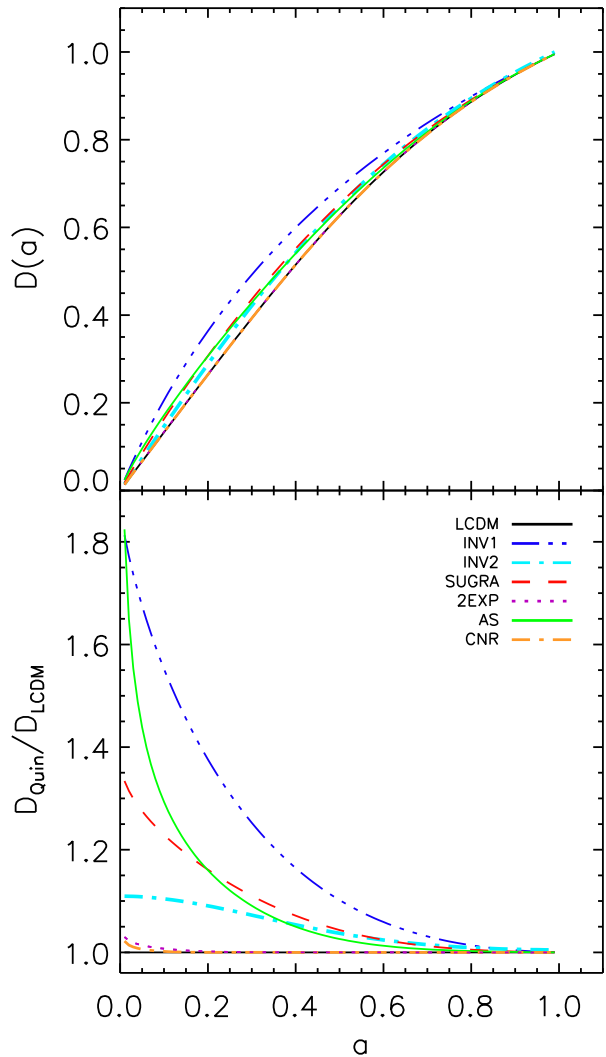}}
\caption{The growth factor as a function of expansion factor.
The upper panel shows the evolution of the  linear growth factor in each quintessence model.
 In the lower panel the ratio of the
growth factor in the quintessence models  compared to $\Lambda$CDM is plotted. The growth factor in each case has been normalised to unity today. }\label{growthfactor}
\end{figure}

\begin{figure}
{\epsfxsize=8.truecm
\epsfbox[96 363 453 701]{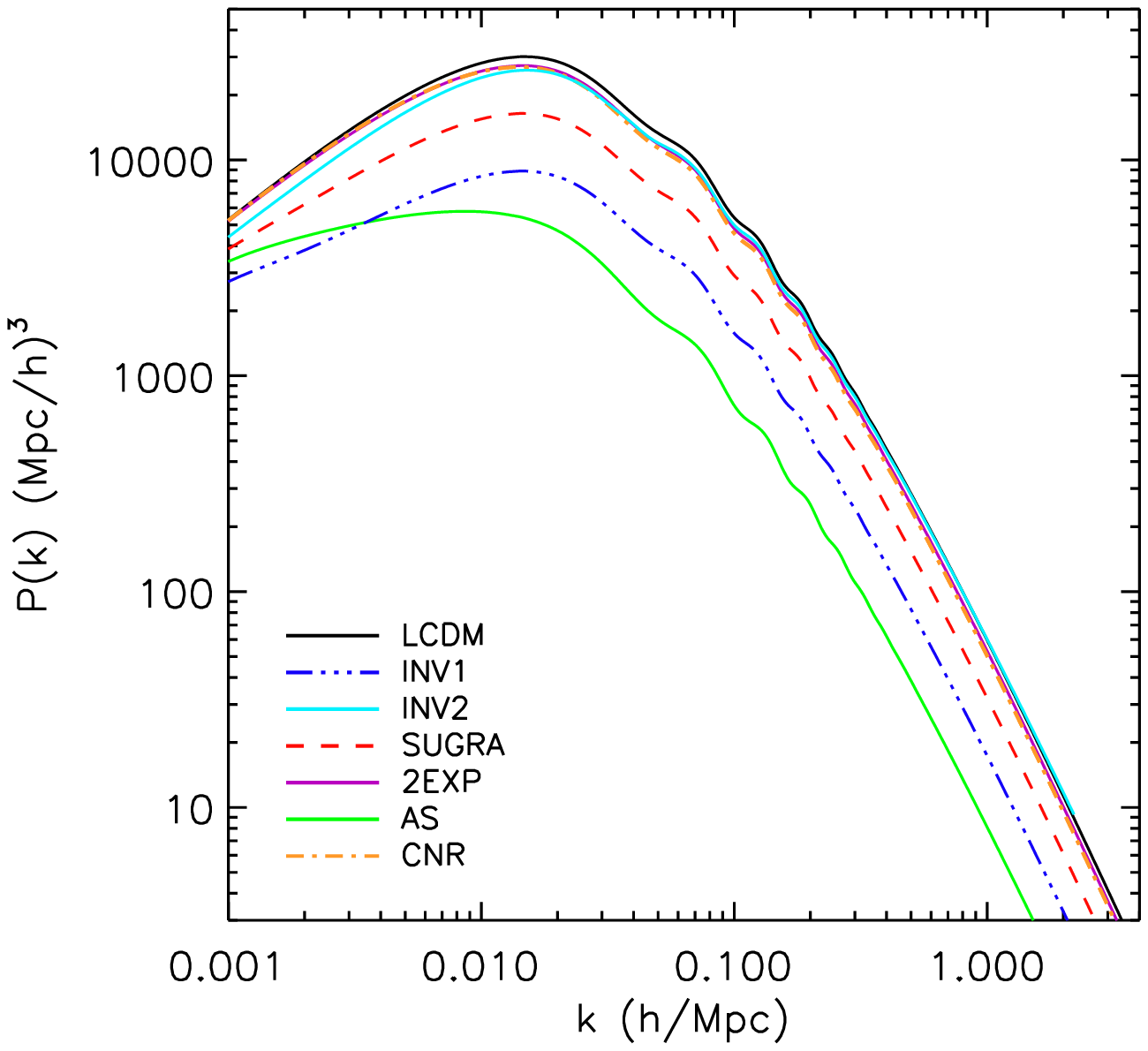}}
 \caption{Linear theory power spectra at $z= 0$ for dynamical dark energy
quintessence models and $\Lambda$CDM.
In this plot, the spectra are normalised to CMB fluctuations (on smaller wavenumbers than are included in the plot).
The presence of a non-negligible dark energy density fraction at
 early times causes a scale independent suppression of growth for scales $k>k_{\tiny \mbox{eq}}$ where $k_{\tiny \mbox{eq}}$
is the wavenumber corresponding to the horizon scale at matter radiation equality and a scale dependent suppression
at $k<k_{\tiny \mbox{eq}}$. Models with high $\Omega_{\tiny\mbox{DE}}$ at the
last scattering surface have a lower   $\sigma_8$ today compared to $\Lambda$CDM if normalised to CMB fluctuations.}
 \label{rawpk}
\end{figure}

\subsection{The expected impact of dark energy on structure formation \label {2.3}}

The growth of structure is sensitive to the amount of dark energy, as this changes the rate of expansion of the Universe.
As a result, a  quintessence model with a varying equation of state could
display different large scale structure from a $\Lambda$CDM model.
Varying the  equation of state will result in different amounts of dark energy at different times. It has been shown that
models
with a larger density of dark energy at high redshift than $\Lambda$CDM  have more developed large scale structure at early times, when
normalised to the same $\sigma_8$ today \citep{Grossi:2008xh,Francis:2008md}.

When the dark matter perturbations are small and  the density contrast $\delta(\vec{x},t) \ll 1$,
the expression for the power spectrum as a function of time, $P(k,t)$, is separable as
\begin{equation}
P(k,t) = \frac{D(t)^2}{D(t_0)^2}\,P(k,t_0),
\end{equation}
where $D(t_0)$ is the linear growth factor at the present epoch.
The normalised growth factor $G = D/a$ obeys the following evolution equation \citep{Linder:2003dr},
\begin{equation}
\label{GF}
G'' + \left( \frac{7}{2} - \frac{3}{2}\,\frac{w(a)}{1+X(a)}\right) \frac{G'}{a} + \frac{3}{2}\frac{1-w(a)}{1+X(a)}\, \frac{G}{a^2} = 0 \, ,
\end{equation}
where
\begin{equation}
X(a) = \frac{\Omega_{\rm m}}
{1-\Omega_{\rm m}}
\,e^{-3\, \int_a^1 \, {\rm d} \mbox{\tiny ln}a' w(a')} \,,
\end{equation}
and  $w(a)$ is the dynamical dark energy equation of state.
The linear growth factor for each quintessence model is plotted in Fig. \ref{growthfactor}.
In Section \ref{psI}, we present the simulation results for each quintessence model
where the initial conditions  were generated using a $\Lambda$CDM linear theory power spectrum and
the background cosmological parameters are the best fit values assuming a $\Lambda$CDM cosmology (Stage I).
The difference
between the simulations is the result of having a  different linear growth rate for the dark matter perturbations.

The presence of small but appreciable amounts of dark energy at early times also modifies the growth
rate of fluctuations from that expected in a matter dominated universe and hence changes the shape of the linear theory $P(k)$ from the $\Lambda$CDM prediction.
The quintessence scalar field can contribute at most a small fraction of the total energy density at early redshifts. Constraints on this amount come from
big bang nucleosynthesis as well as from CMB measurements.
\citet{Bean:2001wt} found a limit of $\Omega_{\tiny \mbox{DE}}<0.045$ at $a \sim 10^{-6}$ using the
observed abundances of primordial nuclides
and a constraint of $\Omega_{\tiny \mbox{DE}}< 0.39$ during the
radiation domination era, $a \sim 10^{-4}$, from CMB anisotropies.
 \citet{Caldwell:2003vp}
discuss the parameter degeneracies which allow for different amounts of dark energy at early times
leaving the position of the CMB peaks unchanged (see Section \ref{wmap}).
Using the WMAP first year data, \citet{Corasaniti:2004sz} found a limit of $\Omega_{\tiny \mbox{DE}} <0.2$ at $z \sim 10$.
Some recent parametrization dependent constraints on early dark energy models found  the dark energy density parameter to be
$\Omega_{\tiny {\mbox{DE}}} <0.02 $ at the last scattering surface \citep{Xia:2009ys}.
Note that all of the models we consider are consistent with this constraint, except for the AS model (see Fig. \ref{omegade}).

If the dark energy is not a cosmological constant, then there will be
dark energy perturbations present, $\delta_{\varphi}$ whose evolution will affect the dark matter
power spectrum and alter the evolution equation in Eq. \ref{GF} \citep{Ferreira:1997hj,Weller:2003hw}.
As most of the
quintessence models we will consider display a non-negligible contribution to the overall density from dark energy
at early times,
 the matter power spectrum is  affected in two ways \citep{Ferreira:1997hj,Caldwell:2003vp,Doran:2007ep}.
In the matter dominated era, the growing mode solution for dark matter
density perturbations is proportional to the expansion factor, $\delta_{\rm m}
\propto a$, in a
universe without a scalar field component.
In a dark energy model which has appreciable amounts of dark energy at early times, the dark matter growing mode
solution  on subhorizon scales is modified to become
\begin{equation}
\delta_{\rm m} \propto a^{[\sqrt{25 -24 \Omega_{\tiny {\mbox{DE}}}} -1]/4} .
\end{equation}
The growth of  modes
on scales $k >k_{\tiny \mbox{eq}}$, where $k_{\tiny \mbox{eq}}$ is the
wavenumber corresponding to the horizon scale at matter radiation equality, is therefore suppressed relative to the growth expected in a $\Lambda$CDM universe.
For fluctuations with wavenumbers  $k<k_{\tiny \mbox{eq}}$ during the matter dominated epoch,
the suppression takes place after the mode enters the horizon and
the growing mode is reduced relative to a model with $\Omega_{\tiny \mbox{DE}} \simeq 0$.
These two effects are illustrated for a scaling quintessence model in
 \citet{Ferreira:1997hj}, whose Figure 7
shows the evolution of $\delta_{\rm m}$ for two wavenumbers, one that enters the horizon
 around $a_{\tiny \mbox{eq}}$
($k=0.1$Mpc$^{-1}$) and one that comes in during the radiation era ($k=1$Mpc$^{-1}$), in a universe
with $\Omega_{\tiny \mbox{DE}}=0.1$ durning the matter dominated era. There is a clear suppression of growth after
horizon crossing compared to a universe with no scalar field.
The overall result is a scale independent suppression for
subhorizon modes, a scale dependent red tilt ($n_s<1$)
for superhorizon modes and an overall broading of the turnover in the
power spectrum.
This change in the  shape of the turnover in the matter power spectrum can be clearly seen in Fig. \ref{rawpk} for the AS model.
This damping of the growth after horizon crossing will result in a smaller $\sigma_8$ value for the quintessence models compared to $\Lambda$CDM if
normalised to CMB fluctuations (see also \citealt{2004PhRvD..70d1301K}).

We have used the publicly available PPF (Parametrized Post-Friedmann) module for CAMB, \citep{Fang:2008sn}, to generate the linear theory power spectrum.
This module supports a time dependent dark energy equation of state by implementing a
PPF prescription for the dark energy perturbations with a constant sound speed $c_s^2=1$.
Fig. \ref{rawpk} shows the dark matter power spectra at $z=0$ generated by CAMB for each quintessence model and $\Lambda$CDM with the same cosmological
parameters, an initial scalar amplitude of $A_s = 2.14 \times 10^{-9}$
and a spectral index $n_s = 0.96$ \citep{Sanchez:2009jq}. As can be seen in this plot,
models with higher fractional energy densities at early times have a lower $\sigma_8$ today and
a broader turnover in $P(k)$.
In Section \ref{psII} a consistent linear theory power spectrum was used for each quintessence model to 
generate the initial conditions for the simulations (Stage II).

Finally, quintessence dark energy models will not necessarily
agree with observational data when adopting
the cosmological parameters derived assuming a $\Lambda$CDM
cosmology.
 We  consider how the different quintessence models affect various distance scales.
We find the best fit cosmological parameters for each quintessence model
using the  observational
 constraints on distances such as the measurements of the angular diameter distance and sound horizon at the last scattering surface
from the cosmic microwave background.
The method and data sets used are given in Appendix \ref{A} and the corresponding simulation results which use a consistent linear theory power spectrum for each model together with the best fit cosmological parameters are presented in Section \ref{wmap} (Stage III).

\section{Simulation Details}

We will determine the impact of quintessence dark energy 
on the growth of cosmological structures  through a series of large N-body simulations.
These simulations were carried out at the Institute of Computational Cosmology using a memory efficient version of the  TreePM
  code  {\tt Gadget-2}, called {\tt L-Gadget-2} \citep{Springel:2005mi}. As our starting point, we consider a $\Lambda$CDM model with 
the following cosmological parameters:  
$\Omega_{\rm m} = 0.26$,
 $\Omega_{\rmn{DE}}=0.74$, $\Omega_{\rm b} = 0.044$,
$h = 0.715$ and a spectral tilt of $n_{\mbox{s}} =0.96$ \citep{Sanchez:2009jq}. 
The  linear theory rms fluctuation 
in spheres of radius 8 $h^{-1}$ Mpc is set to be  $\sigma_8 = 0.8$. 
For each of the quintessence models the parametrization 
for the dark energy equation of state given in Eq. \ref{weq} was used. 
In the first stage we fix the cosmological parameters  for all of the quintessence models to those of $\Lambda$CDM.
As a result, some of the scalar field models do not match observational constraints on
the sound horizon at last scattering or the angular diameter distance. We shall discuss this further in 
Section \ref{wmap} using the results given in Appendix \ref{A}.

The simulations use $N=646^3 \sim 269 \times 10^6$ particles to represent the dark matter in a  computational box of 
comoving length $1500 h^{-1}$Mpc. 
 We chose a  comoving softening length of  $\epsilon = 50 h^{-1}$kpc. 
The particle mass in the simulation is $9.02 \times 10^{11}  h^{-1}
M_{\sun}$ with a mean interparticle separation of
$r \sim 2.3$ $h^{-1}$Mpc.
The initial conditions of the particle load were set up with a 
 glass configuration of particles. 
This arrangement is obtained by evolving a random distribution of particles
 with the sign of the gravitational force reversed \citep{1994RvMA....7..255W,Baugh:1995hv}.
The particles are perturbed from the glass using the Zeldovich approximation which
can induce small scale transients in the measured power spectrum. These transients die away after $\simeq$10 expansion factors from the starting redshift  
\citep{Smith:2002dz}.
In order to limit the effects of the initial displacement scheme
 we chose a starting redshift of $z=200$. 

The linear theory power spectrum used to generate the initial 
conditions was created using the CAMB package of \citet{Lewis:2002ah}.  
 In the first stage of our calculations, presented in
Section \ref{psI}, the  linear theory power spectrum used to set up the initial conditions in the quintessence models was the same as 
 $\Lambda$CDM.
For the purpose of computing the shape of $P(k)$ in Stage I, we have assumed that the ratio of dark energy density to the critical density at
the last scattering surface ($z_{\mbox{lss}} \sim 1000$) is negligible  and have ignored any clustering of the
scalar field dark energy.
 In Section \ref{psII}, the linear theory $P(k)$ is generated for each quintessence 
model using a modified version of CAMB which incorporates 
the influence of dark energy on dark matter clustering at early times. 
 In each model the power spectra at redshift zero have been normalised to have 
$\sigma_8 = 0.8$. Using the linear growth factor for each dark energy model, the 
linear theory $P(k)$ was then evolved backwards to the 
starting redshift of $z=200$ in order to generate the initial 
conditions for {\tt L-Gadget-2}. 
Snapshot outputs of the dark matter distribution as well as the group catalogues were made at redshifts 5, 3, 2.5, 2, 1.5, 1, 0.75, 0.5, 0.25 and 0.
The simulation code {\tt L-Gadget-2} has an inbuilt friends-of-friends (FOF) group finder which was applied
to produce  group catalogues of dark matter particles with 10 or more particles. A linking length of 0.2 times the mean interparticle separation was used in the group finder.

 We investigate
 gravitational collapse in the six
quintessence models  listed in Table 1
by comparing the evolution of the power spectrum at
 various redshifts.
The power spectrum was computed by assigning the particles to a mesh using the cloud in cell (CIC) assignment scheme \citep{1981csup.book.....H} and then
performing a fast Fourier transform on the density field.
To restore the resolution of the true density field this assignment scheme is corrected for by performing an approximate de-convolution  \citep{1991ApJ...375...25B}.

\begin{figure}
{\epsfxsize=8.5truecm
\epsfbox[96 363 453 701]{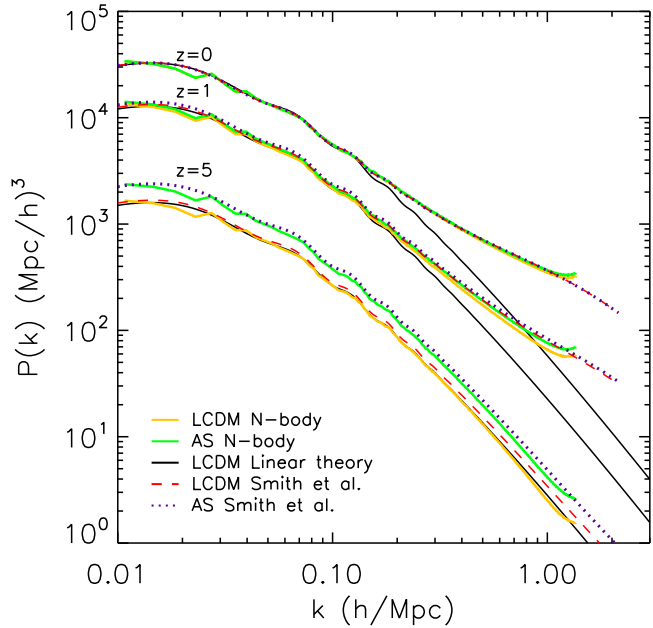}}
 \caption{Power spectra in a $\Lambda$CDM cosmology (orange lines) and  AS quintessence model (green lines) at redshift 0, 1 and 5.
The red dashed lines corresponds to the \citet{Smith:2002dz} analytical expression for the nonlinear $P(k)$ in
$\Lambda$CDM; blue dotted lines show the equivalent for the AS model. The solid black line is the linear theory for $\Lambda$CDM at the corresponding redshift outputs. The \citet{Smith:2002dz} expression for the
AS model has been scaled with the appropriate growth factor for this model at each redshift.}
 \label{linearpk}
\end{figure}
\begin{figure*}  \center
{\epsfxsize=12.truecm
 \epsfbox[82 360 424 708]{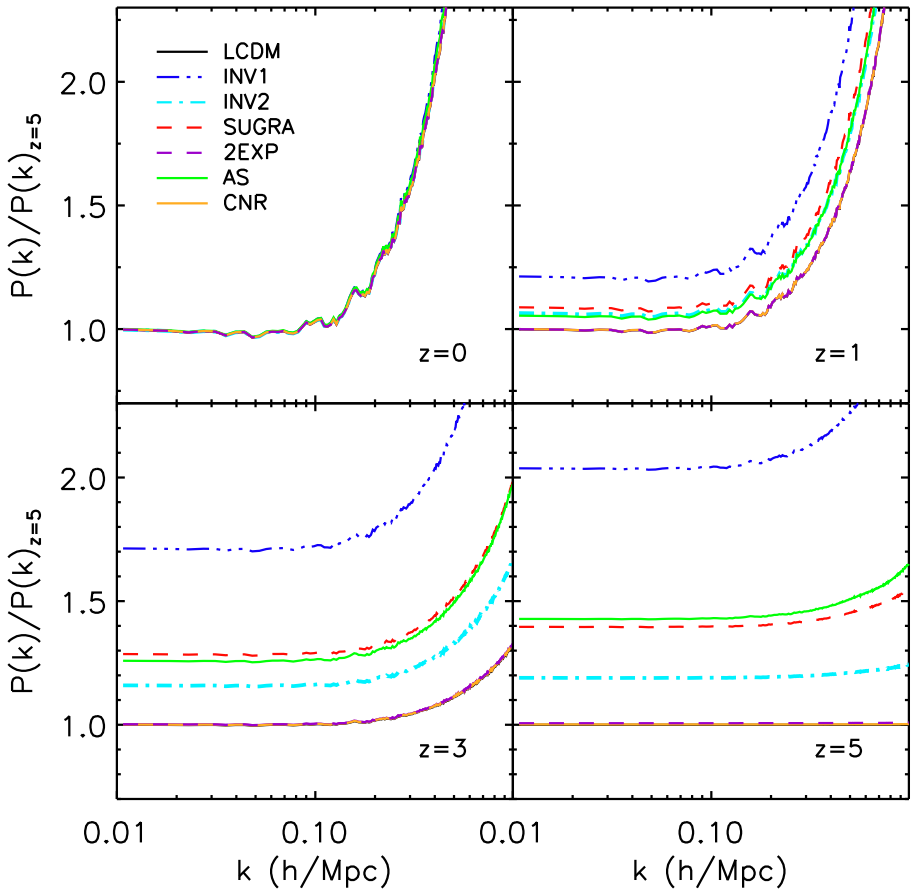}}
\caption{The nonlinear growth of the power spectra in the various quintessence models as indicated by the key in the top left panel.
Each panel shows a different
redshift. The power spectra in each case have been
divided by the $\Lambda$CDM power spectrum at redshift 5 scaled to take out the difference between the $\Lambda$CDM growth factor at $z=5$ and the redshift plotted in the panel.
 This
 removes the sampling variance due to the finite box size and highlights the enhanced nonlinear growth found in
quintessence cosmologies compared to $\Lambda$CDM.
A deviation of the power ratio from unity therefore indicates a difference in $P(k)$ from the linear perturbation theory of $\Lambda$CDM.}
\label{pk}
 \end{figure*}

\begin{figure}
{\epsfxsize=8.5truecm
\epsfbox[96 363 453 701]{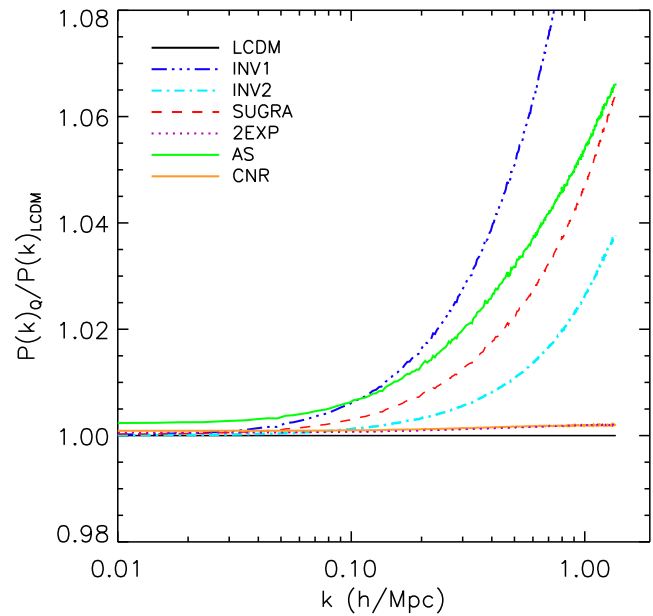}}
 \caption{Ratio of power spectra output from the simulations in the six quintessence models compared to the nonlinear $\Lambda$CDM $P(k)$ at redshift 0.
Note the expanded scale on the y-axis.
As expected, the 2{\sc{EXP}} and CNR models show no difference
from $\Lambda$CDM while
the difference in the  INV1,  INV2,  SUGRA and AS
 models is under 10\% for wavenumbers $k<1 h$Mpc$^{-1}$.   }
\label{z0pk}
\end{figure}

\section{Results}
In the following sections we present the power spectrum predictions from the three stages of simulations carried out as described in Section 2.3.
 The bottom line results are presented in Section 4.3, in which we compare power spectra in $\Lambda$CDM with a subset of dark energy models which also 
pass the currently available observational constraints. The reader pressed for time may wish to skip directly to this section.
Sections \ref{psI} and \ref{psII} show intermediate steps away from $\Lambda$CDM towards the consistent dark energy models presented in Section \ref{wmap}, to allow us to 
understand the impact on $P(k)$.
In Section \ref{psI} the Friedmann equation was modified with the quintessence model's equation of state as a function of redshift and 
a $\Lambda$CDM linear theory power spectrum was used to generate the initial conditions for all the simulations (Stage I).
In Section \ref{psII} we use a consistent linear theory power spectrum for each quintessence model 
(Stage II).
In Section \ref{wmap} we constrain a set of cosmological parameters, using CMB, BAO and SN data, for each dark energy model. The final stage
 of simulations use a consistent linear theory power spectrum for each model together with the best fit cosmological parameters 
(Stage III).

\subsection{Stage I : Changing the expansion rate of the Universe \label{psI}}

In this first stage of simulations, the same $\Lambda$CDM initial power spectrum and cosmological parameters were used for all models.
In Fig. \ref{linearpk} we plot the power spectrum at redshifts $z = 0,1,5$ in $\Lambda$CDM (orange lines) and in the  AS model (green lines),
together with the linear theory power spectra
for $\Lambda$CDM (black lines).
The AS model has a linear growth rate that differs from $\Lambda$CDM by $\sim 20\% $ 
at $z=5$. 
We also plot the \citet{Smith:2002dz} \lq Halofit\rq \, empirical fitting function for $\Lambda$CDM and the AS model.
The Halofit function has been incorporated into the CAMB package and this code
 was used to generate the output at various redshifts seen in
Fig. \ref{linearpk}. As this plot shows, the \citet{Smith:2002dz} expression accurately describes the evolution of the power spectrum at redshift 0 in both
models and at earlier times. As the normalisation and linear spectral shape is the same in these two models, Halofit accurately
reproduces the nonlinear power in each model at various redshifts once the appropriate linear growth factor 
for the dark energy model at that redshift is used.
The Smith et al. expression agrees with the simulation output at $z=0$ to within 4\% for $k<1 h$Mpc$^{-1}$ for both the quintessence model 
and $\Lambda$CDM. At higher 
redshifts,  the difference between the simulation output and the Halofit prediction for all the 
models is just 
under 10\% on scales $k<0.3 h$Mpc$^{-1}$ at $z=5$.

To highlight the differences in the power between the different models, we plot in Fig. \ref{pk} the measured power 
divided by the power at $z=5$, after scaling to take into account the difference in the linear theory growth factors for the output redshift and $z=5$, for $\Lambda$CDM. 
This removes the sampling variance from the plotted ratio \citep{1994MNRAS.270..183B}. 
A ratio of unity  in Fig. \ref{pk} would indicate linear growth at the same rate as expected in $\Lambda$CDM.
\begin{figure*}
{\epsfxsize=12.truecm
\epsfbox[71 361 422 703]{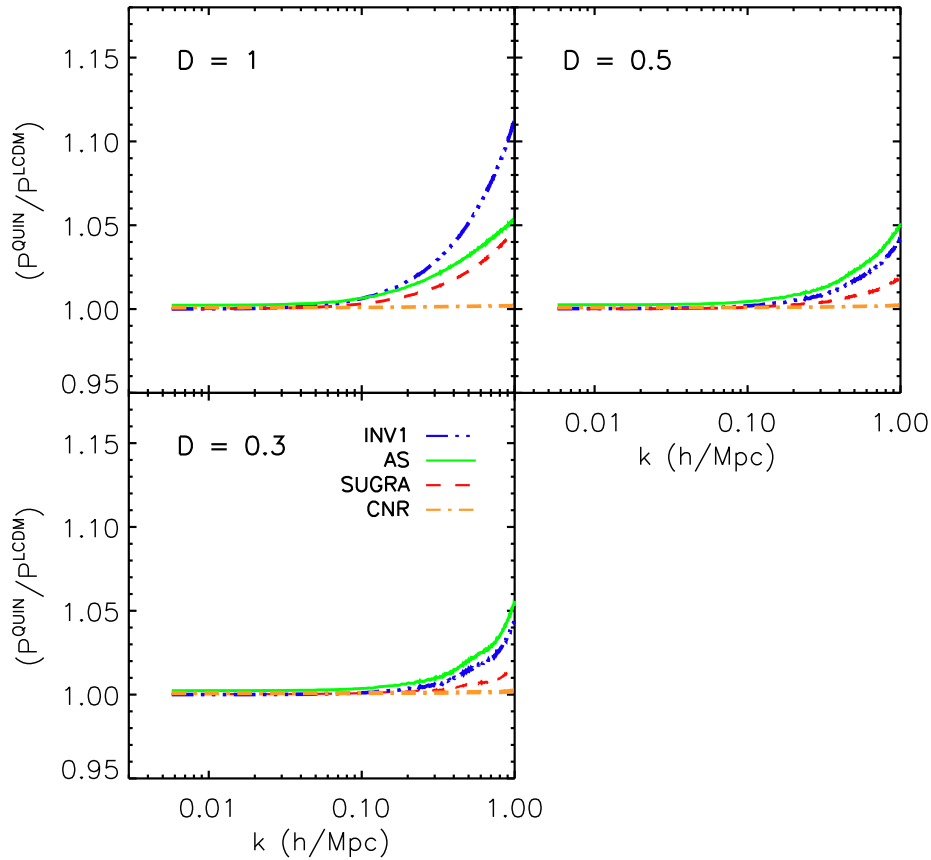}}
 \caption{ The ratio of the quintessence model power spectra to the $\Lambda$CDM
power spectrum output from the simulations at three values of the linear growth factor $D = 1, D = 0.5$
and $D = 0.3$.
Each panel shows the results of this exercise for the AS, CNR, 2EXP and SUGRA quintessence models. The growth factors
correspond to $z=3.4$ ($D=0.3$), $z=1.6$ ($D=0.5$) and $z=0$ ($D=1$) for $\Lambda$CDM.
For each model, the choice of growth factor corresponds to slightly different redshifts, with the biggest difference being
for the INV1 model. A ratio of unity would indicate that the growth factor is the only ingredient needed
to predict the power spectrum in the different quintessence models. Note the expanded scale on the y axis.
}
\label{scaledgrowth}
\end{figure*}

Fig. \ref{pk} shows four epochs in the evolution of 
the power spectrum for all 
of the quintessence models and $\Lambda$CDM. 
The black line in the plot shows the $P(k)$ ratio for $\Lambda$CDM (note the yellow curve for the 
CNR model is overplotted). 
Non-linear growth can be seen as an increase in the power ratio on small scales, 
$k>0.3h$Mpc$^{-1}$ at $z=3$ and $k>0.1h$Mpc$^{-1}$ at $z=0$. 
 Four of the quintessence models (INV1, INV2, SUGRA and AS) differ significantly from $\Lambda$CDM for $z>0$. 
These models  show   advanced
 structure formation i.e. more power than $\Lambda$CDM, and a large increase in the 
 amount of nonlinear growth.
All models are normalised to have $\sigma_8 =  0.8 $ today 
and as a result all the power spectra are very similar at redshift zero in Fig. \ref{pk}.  There are  actually small differences between  the quintessence models 
at $z=0$ as seen on the expanded scale in  Fig. \ref{z0pk}. This increase in nonlinear power at small scales in the 
quintessence models is 
due to the different growth histories.

The power spectra predicted in the 2EXP and CNR models show minor departures from that in the $\Lambda$CDM cosmology. 
This is expected as Figs. \ref{w} and \ref{omegade}
show the equations of state and  the dark energy densities in these two models are the same as $\Lambda$CDM at low redshifts 
and all three simulations began from identical initial conditions. It could be 
possible to distinguish these two models from the 
concordance cosmology at higher redshifts if we do not ignore the dark energy perturbations or changes in the growth factor which alter  the form of the linear theory 
power spectrum. We shall discuss this more in the next stage of our simulations in Section \ref{psII}. 

Finally, we investigate  if the enhanced growth in the power spectrum seen in Fig. \ref{pk} in the quintessence models is due solely 
to the different linear growth rates at a given redshift in the models. 
In order to test this idea, the power spectrum in a quintessence model and $\Lambda$CDM 
are compared not at the same redshift but at the same linear growth factor \footnote{We thank S. D. M. White for this suggestion.}. 
As the growth rates in some of the 
quintessence models are very different from that in the standard $\Lambda$CDM cosmology, the power spectra
required from the simulation will be at different output redshift in this comparison.
For example, the normalised linear growth factor is $D = 0.5$ at a redshift of $z=1.58$ in a $\Lambda$CDM model and has the same value 
at  $z = 1.82$ in the SUGRA model, at $z = 1.75$ in the AS model and at 
$z=2.25$ in the INV1 quintessence model. 
In Fig. \ref{scaledgrowth} we show the power spectrum of  simulation outputs from the INV1, AS, SUGRA and CNR models divided 
by the power spectrum output in $\Lambda$CDM at the same linear growth rate. We ran the simulations taking three additional redshift
 outputs where the linear growth rate had values of $D = 1, D = 0.5$ and 
$D = 0.3$. 
It is clear from Fig. \ref{scaledgrowth} that scaling the power spectrum in this way can explain the enhanced linear and most of the excess nonlinear growth seen in 
Fig. \ref{pk} for scales $k<0.1 h$Mpc$^{-1}$. 
For example, in the INV1 model the enhanced nonlinear growth,
on scales $k \sim 0.3h$Mpc$^{-1}$ at fixed $D=0.3$, 
differs from $\Lambda$CDM by at most 5\% in Fig. 
\ref{scaledgrowth} as opposed to at most 30\% at $z=5$ in Fig. \ref{pk}.
At earlier redshifts when the linear growth rate is $D=0.3$, the nonlinear growth 
in the quintessence models agrees with
$\Lambda$CDM on  smaller wavenumbers $k<0.3 h$Mpc$^{-1}$. As in Fig. \ref{pk}, the CNR model shows no difference from $\Lambda$CDM when plotted in this way. 

Note in Fig. \ref{scaledgrowth}  the INV1 model has less nonlinear growth at $D=0.3$ and $D=0.5$ compared to the AS  model.
The AS and SUGRA models have a growth
rate of
D=0.5 at lower redshifts compared to the INV1 model and so are at a later stage in their growth history.
The INV1 model has a growth rate of $D=0.5$
at $z=2.25$ whereas for the AS model this occurs at $z=1.75$ and at $1.82$ for the SUGRA model.
The reason for the success of this simple model - matching the growth factor to predict the clustering - 
can be traced to the universality of the mass function, which we discuss in Section \ref{mf}.
In this Stage I calculation, the models have the same mass function when plotted at the epoch
corresponding to a common growth factor. This means that the two-halo contribution to the clustering is therefore the same.
Can this simple halo picture of the clustering also explain the clustering on small scales (high $k$)? Although the abundance of haloes in the 
models is the same at the epochs corresponding to a given value of the growth factor, the concentrations of the haloes will not be the same.
In cosmologies where the haloes formed at a higher redshift (i.e. roughly the redshift corresponding to a particular value of $D$), 
one would expect these haloes to have higher concentrations than their counterparts in the other models \citep{2001ApJ...554..114E}. 
A higher concentration would be expected to yield stronger 
nonlinear clustering and hence more power at high $k$ in Fig. \ref{scaledgrowth}. Unfortunately our simulations do not have the resolution to 
probe the required range of wavenumbers to uncover this behaviour. The ratios plotted in Fig. \ref{scaledgrowth}  stop at wavenumbers 
approximately equivalent to the collapsed radius of a massive halo.

Hence, it seems that scaling the power spectrum using the linear growth rate can be used to predict the linear growth in the quintessence dark energy 
simulations and can reproduce 
some of the nonlinear growth at early 
redshifts. In Fig. \ref{scaledgrowth} there are still some differences in the small scale growth in quintessence models compared to $\Lambda$CDM which cannot be explained
by the different linear growth rates.
We find that nonlinear evolution is not just a function
of the current value of the linear growth rate but also depends on its history through the  evolution of the coupling between long and
short-wavelength modes.

\subsection{Stage II: Use of a self-consistent linear theory $P(k)$ \label{psII}}

\begin{figure}
{\epsfxsize=8.5truecm
\epsfbox[96 363 453 701]{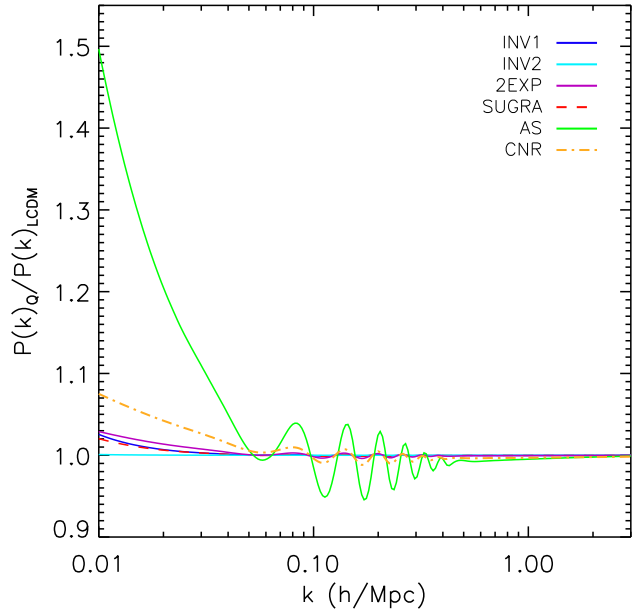}}
 \caption{Ratio of linear theory power spectra  for quintessence models shown in Fig. \ref{rawpk}
 to that in $\Lambda$CDM. In this plot each $P(k)$ has been normalised so that  $\sigma_8 = 0.8$ today; this is
the normalisation used in our simulations.}
 \label{PlotPk}
\end{figure}

\begin{figure*}
\center
{\epsfxsize=16.truecm
\epsfbox[71 366 424 701]{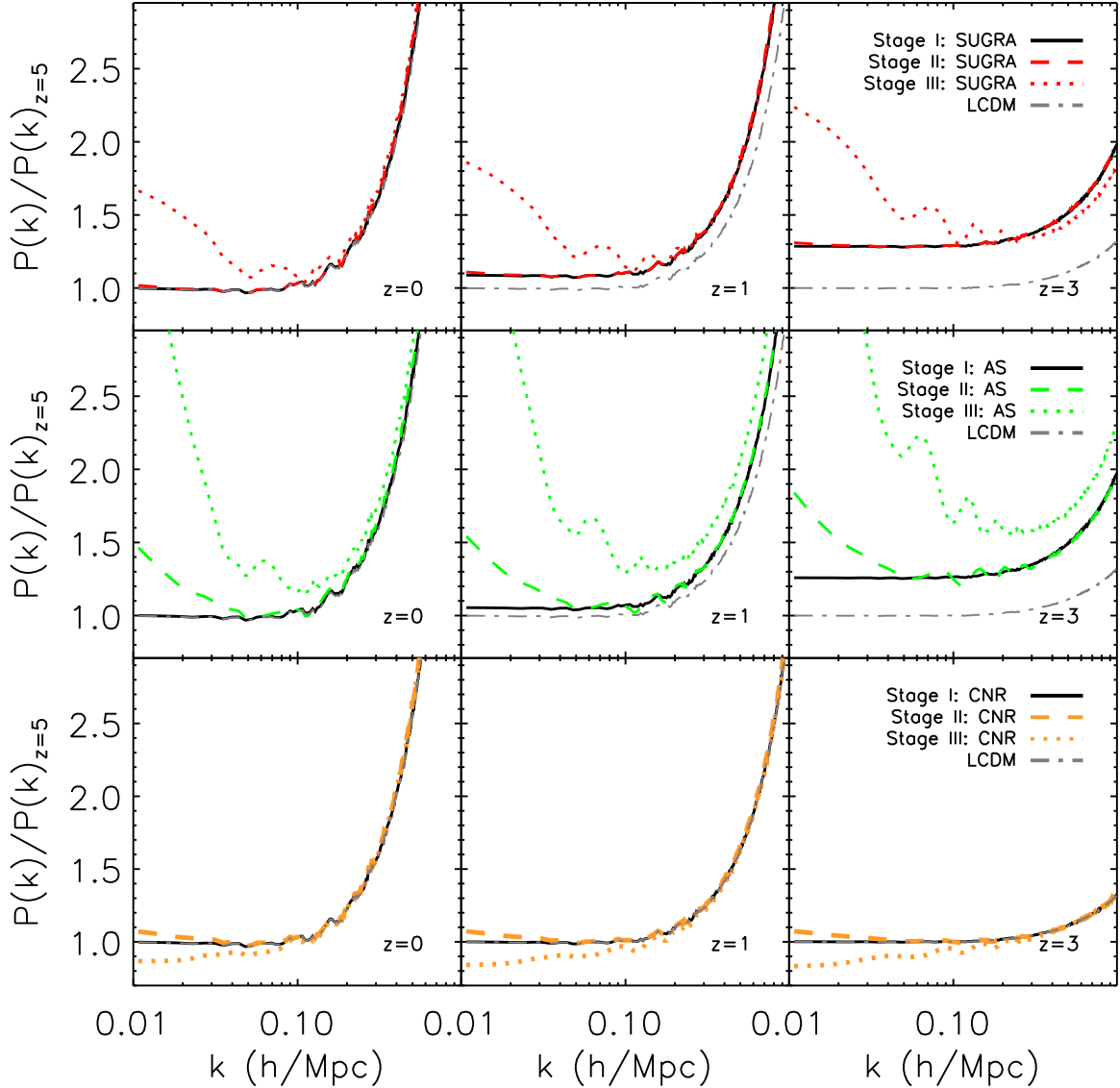}}
 \caption{Ratios of power spectra for the SUGRA (first row), AS (second row) and CNR (third row)
quintessence model compared to $\Lambda$CDM from the 3 stages of simulations in this paper.
The plot shows the growth in the quintessence models using $\Lambda$CDM
linear theory $P(k)$ in the initial conditions in black (Stage I)
and using
a self consistent linear theory $P(k)$ for each quintessence model
 (dashed colored line) (Stage II). The dotted lines shows the $P(k)$ ratio from the simulation for the quintessence models using the
best fit parameters in Table A3 (Stage III).
The power spectra in each case have been
divided by the $\Lambda$CDM power spectrum at redshift 5 with appropriate scaling of $\Lambda$CDM
growth factors. The linear theory
power spectra in each case has been normalised to $\sigma_8 =0.8$.
}
\label{MOD}
\end{figure*}

We have run the simulations presented in the previous section again but this time 
using the appropriate linear theory $P(k)$ for each model (shown in Fig. \ref{rawpk})  normalised to  
$\sigma_8 =0.8 $ today (Stage II). 
After normalising the power spectra in this way, the difference between the quintessence models $P(k)$ and $\Lambda$CDM can be 
seen in Fig. \ref{PlotPk}. 
The INV2 model was not included in this set of simulations as there is a  negligible difference in the linear theory power 
spectrum from $\Lambda$CDM. Note  \citet{Francis:2008md} 
also generate the linear theory power spectrum for  \lq early dark energy\rq \, models and normalise all $P(k)$ to have the 
same $\sigma_8$ today. 
 \citet{Francis:2008md} make an equivalent plot to Fig. \ref{PlotPk} but 
 find a decrease in this ratio with decreasing scale ($k>0.2 h$Mpc$^{-1}$), using the parametrization for early dark energy proposed by \citet{Doran:2006kp}, 
in contrast to the ratio of unity we find on 
small scales in Fig. \ref{PlotPk}. 
This difference is due to the different parametrizations used for the dark energy equation of state, as a ratio of unity is obtained on small scales for the same 
\lq early dark energy\rq \, model using the parametrization suggested by \citet{Wetterich:2004pv} (M. Francis, private communication).

In the first row of Fig. \ref{MOD} we plot 
the power spectrum for the Stage II SUGRA model at $z= 0$, 1, and 3  divided by  the simulation output in $\Lambda$CDM  at $z= 5$ as in
Fig. \ref{pk} (red dashed lines). 
The result from Fig. \ref{pk}, Stage I SUGRA,  is also plotted here to highlight how changing the spectral shape affects the nonlinear growth in 
the simulations. On large scales the growth is not modified by the altered spectral shape. 
The growth of perturbations on small scales in the simulation is
affected by the modified linear theory used in the initial conditions. 
Normalising the power spectra to $\sigma_8 =0.8$ results in more power on 
large scales in the quintessence models compared to $\Lambda$CDM, as can be seen in Fig. \ref{PlotPk}. This enhanced large scale power 
couples to the power on smaller scales and results in a small  increase in the nonlinear power spectrum for $k>0.1 h$Mpc$^{-1}$ in 
the Stage II SUGRA simulation compared to the one using $\Lambda$CDM linear theory $P(k)$
 in Stage I.

In the second row of Fig. \ref{MOD} we plot
the power spectrum for the Stage II AS model as green dashed lines at $z= 0$, 1, and 3  divided by  the simulation output in $\Lambda$CDM  at $z= 5$ as in
Fig. \ref{pk}.
The growth of dark matter
perturbations is greatly suppressed in the AS model due to the large
fractional dark energy density at high redshifts.
 After fixing $\sigma_8 =0.8$, there is more power
on large scales in the AS model compared to $\Lambda$CDM. 
As in the first row of Fig. \ref{MOD} there is a small increase in nonlinear power for the AS model in Stage II.
Although the excess large scale
power is significantly larger than in the SUGRA model case, it does not result in 
more nonlinear power on small scales through mode coupling,
 as can be seen in the panels in the second row in Fig. \ref{MOD}.
The linear theory power spectrum for these quintessence models has a scale dependent 
red tilt on large scales which shifts the position of the BAO peaks which is the origin of the oscillation apparent in the second row of Fig. \ref{MOD} at $z=3$. 
The difference in BAO peak positions is very prominent when we plot the ratio of the power spectrum in the AS model to the $\Lambda$CDM power spectrum and can be clearly seen in Fig. \ref{MOD}.

\subsection{Stage III: Consistency with observational data \label{wmap}}

In this section we present the power spectra results in $\Lambda$CDM and a subset of the dark energy models, measured from 
simulations which use a consistent linear theory power spectrum for each model together with the best fit cosmological parameters.
We have simulated the SUGRA, AS and CNR models using the best fit cosmological parameters from Table A3 and
the linear theory power spectrum specific to each model as discussed in Section \ref{2.3}.
 We chose to simulate these three models following the analysis and 
results of Sections \ref{psI}, \ref{psII}  and Appendix \ref{A}.
Any of the dark energy models listed in Section 2.2 which showed similar results in Section \ref{psII} to $\Lambda$CDM and similar 
cosmological parameters in Appendix \ref{A}  have not been simulated again. 

Table A3 in Appendix \ref{A} shows the best fit values for $\Omega_{\rm m}h^2$, $\Omega_{\rm b}h^2$ and $H_0$ for each quintessence model, found by minimising 
$\chi^2_{\tiny \mbox{total}}=\chi^2_{\tiny \mbox{WMAP+SN+BAO}}$.
The SUGRA, AS and CNR models  had the biggest improvement in the agreement with observational constraints,
on allowing $\Omega_{\rm m}h^2$, $\Omega_{\rm b}h^2$ and $H_0$ to vary. 
 The results for the 
SUGRA, AS and the CNR model are shown as dotted coloured lines in Fig. \ref{MOD} 
and are referred to as Stage III in the legend to distinguish them from the results of Sections \ref{psI} and \ref{psII} which are also plotted. 
In each row we show the simulation outputs at $z=0, 1$ and 3.
The simulation results 
for each quintessence model uses the models linear theory and the best fit parameters 
from Table A3.
Using the best fit parameters for each model together with the correct linear theory changes the growth of  structure in the simulation.

\begin{figure*}
\center
{\epsfxsize=18.truecm
\epsfbox[82 398 541 703]{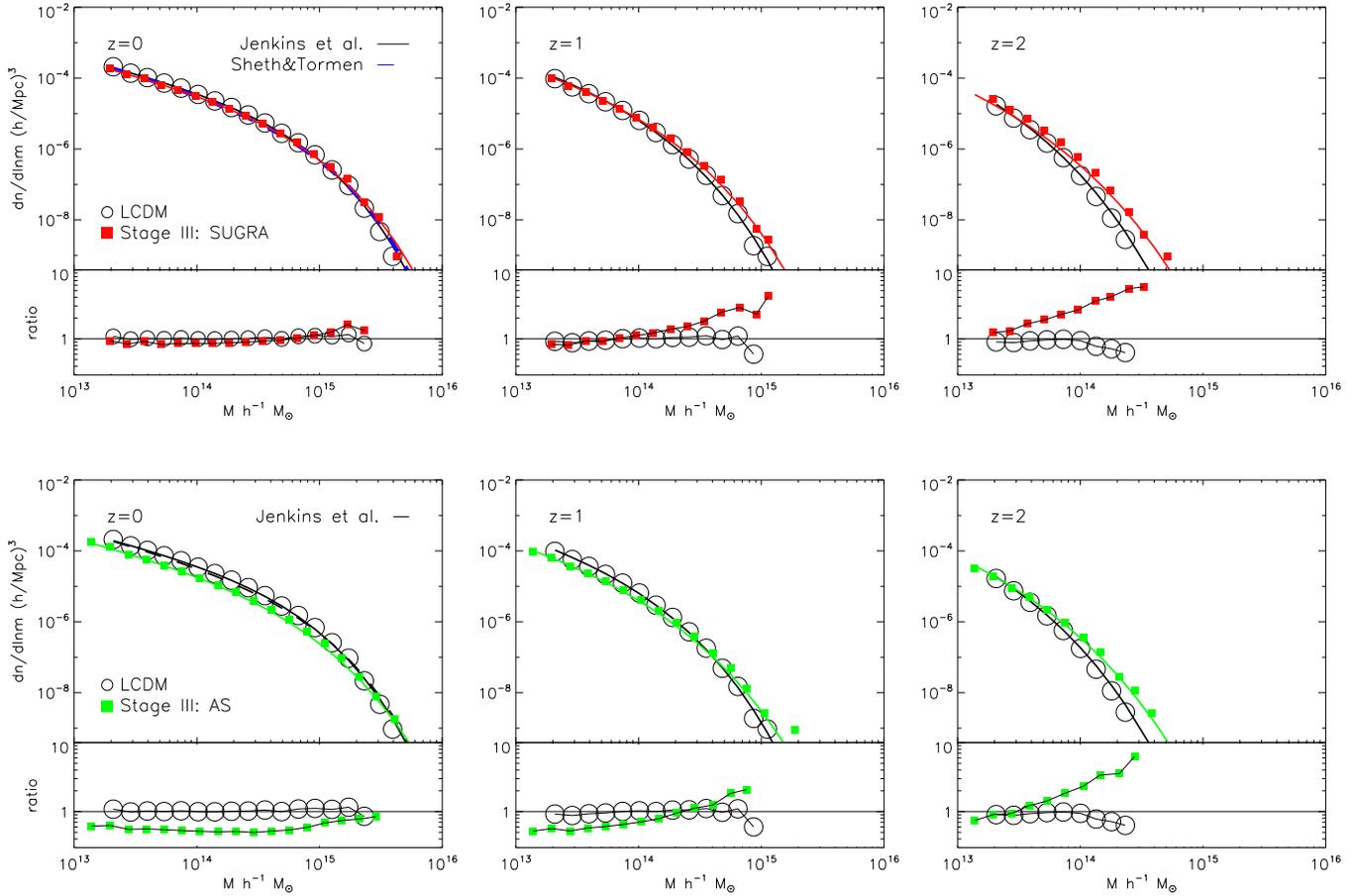}}
 \caption{Dark matter halo mass functions for the  SUGRA (first row) and AS (second row)
quintessence models compared with that in $\Lambda$CDM from the  Stage III simulations at
 $z=0$, 1 and 2.
The mass function in $\Lambda$CDM is shown as open black circles throughout this plot.
In the first row
the red filled squares show the mass function from the simulation for the SUGRA model using the best fit parameters in Table A3 (Stage III).
Underneath each panel in  the first row we plot the log of the ratio between
 the measured mass function for  $\Lambda$CDM (open black circles) and Stage III SUGRA
(red squares) and
the Jenkins mass
function for $\Lambda$CDM.
In the second row the green filled squares show the mass function from the simulation for the AS model using the best fit parameters in Table A3 (Stage III).
For the  AS Stage III  simulation, $\Omega_{\rm m}
h^2 = 0.086$, giving rise to a
change in the spectral shape of the linear theory power spectrum.
As a result, there are fewer low mass halos and a similar number of high mass haloes at $z=0$ compared to $\Lambda$CDM ($\Omega_{\rm m}
 h^2 =0.1334$).
The difference between the Jenkins et al. mass
function for $\Lambda$CDM and the measured mass function for $\Lambda$CDM (open black circles) and Stage III AS
(green squares) is plotted underneath each panel in the second row.
The black horizontal line indicates a ratio of unity in the ratio plots.
In the first and second rows the solid black (red/green) lines are the predicted abundances in the $\Lambda$CDM (SUGRA/AS) model using the Jenkins et al.  fitting function at various redshifts.
In the top left panel, for reference,
we have also plotted the Sheth \& Tormen mass function (blue dashed line) for $\Lambda$CDM.
}\label{J01}
\end{figure*}

In Fig. \ref{MOD} the measured power spectrum for each model is divided by the power for $\Lambda$CDM at $z=5$ which has been scaled using the 
difference in the linear growth factor between $z=5$ and the redshift shown.  Plotting the ratio in this way  highlights
 the differences in growth between the quintessence models and $\Lambda$CDM as well as removing sampling variance.

The measured power for the SUGRA model is plotted in the first row in Fig. \ref{MOD}. 
The power spectra have all been normalised to $\sigma_8 =0.8$ resulting in a large increase in the large scale power ($k<0.1 h$Mpc$^{-1}$)
seen in Fig. \ref{MOD} compared to $\Lambda$CDM.
There is a large increase in the linear and nonlinear growth in this model at $z>0$ (dotted red line) compared to $\Lambda$CDM (dot-dashed grey line).
 The second row in Fig \ref{MOD} shows there 
is a significant enhancement in the
growth in the AS power spectrum measured compared to $\Lambda$CDM for $z<3$.
The power measured from the simulations of the CNR model are plotted in the third row of Fig. \ref{MOD}. We find there is a 
small reduction in the amount of linear and nonlinear growth in this model compared to $\Lambda$CDM.

In Fig. \ref{MOD}  we also plot the simulation results for these three models 
from Section 4.1 (Stage I), where $\Lambda$CDM linear theory was used in the initial conditions,  
 (black lines).
The dashed coloured lines show the simulation results from Section 4.2 (Stage II), where the quintessence model linear theory was used.
The SUGRA power spectrum measured in Stage III has less nonlinear growth at high redshifts compared to the SUGRA $P(k)$ from Stage I or II
due to changes in the  spectral shape. 
The measured power for the AS  model using the best fit parameters (Stage III) 
shows enhanced growth on all scales compared to the power for the AS model in Stage I (using $\Lambda$CDM parameters and linear theory $P(k)$) or Stage II (using $\Lambda$CDM parameters).

These results show the importance of each of the three stages 
in building up a complete picture of a quintessence dark energy model.
Models whose equation of state is very different from $\Lambda$CDM at low redshifts, 
for example the SUGRA and the AS model, show enhanced 
nonlinear growth today compared to $\Lambda$CDM. 
Models whose equation of state is very different to $\Lambda$CDM only at early times, 
for example the CNR model, will show no difference in 
the nonlinear growth of structure if we use the $\Lambda$CDM spectral shape 
(Stage I). In Stage II and III the shape of the power spectrum in the CNR model has changed and is very different to  $\Lambda$CDM on large scales as can be seen in 
Fig. \ref{MOD}.  Using the best fit cosmological parameters 
for this model we find a very small reduction ($<2$\%) 
in the nonlinear growth at $z=0$ compared to $\Lambda$CDM.

\subsection{Mass function of dark matter haloes \label{mf}}

\begin{figure*}
\center
{\epsfxsize=16.truecm
\epsfbox[83 491 540 685]{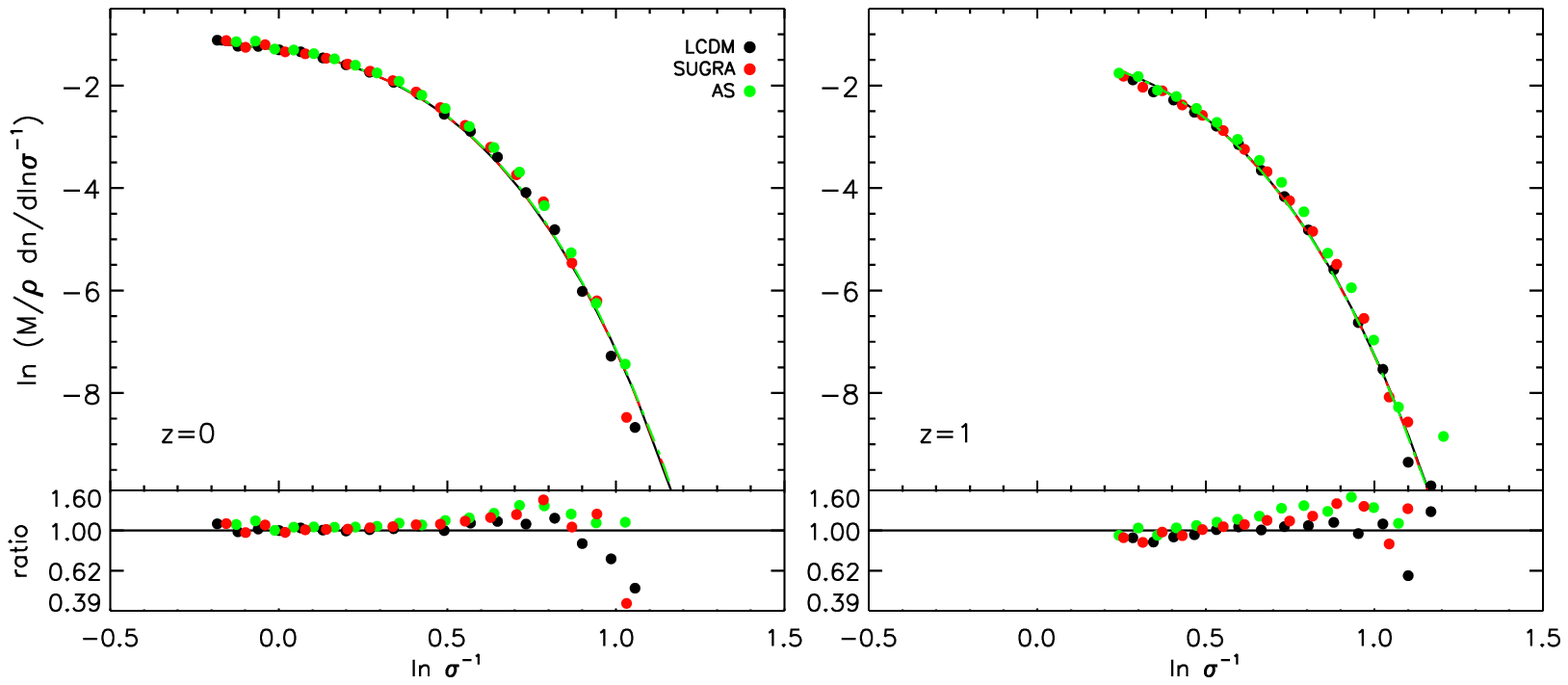}}
 \caption{
The halo mass function for the SUGRA and AS model and $\Lambda$CDM at $z=0$
and 1  compared to the \citet{Jenkins:2000bv} analytic fit. The
Jenkins et al. mass function is plotted as solid black (red/green) lines for  $\Lambda$CDM (SUGRA/AS).
Underneath each panel
the ratio of the mass function
measured from the simulation and the Jenkins et al.
mass function is plotted for all models.
Note a logarithmic scale  is used on the y axis in the ratio plots.
}\label{universal}
\end{figure*}

In this section we present the mass function of dark matter haloes in the quintessence models using
the three stages of simulations discussed in Sections \ref{psI}, \ref{psII} and \ref{wmap}.

\citet{Press:1973iz} (hereafter P-S) proposed an analytical expression for the
abundance of collapsed objects with mass $M$  in the range
$M $ to $ M +dM$ at redshift $z$,
based on the spherical collapse model in which 
a perturbation can be associated with a virialised object at $z=z'$,
 if its density contrast, extrapolated to $z=z'$ using linear theory,
 exceeds some threshold value, $\delta_c$, the critical linear density contrast.
It has been shown that
the P-S approach fails to reproduce the abundance of haloes found in simulations,
 overpredicting the number of haloes below the characteristic mass $M_*$
and underpredicting the abundance in the high mass tail \citep{1988MNRAS.230P...5E,White:1992ri,Lacey:1994su,Eke:1996ds,Governato:1998pf}.

It is thought that the main cause of this discrepancy is the spherical
collapse approximation, as the perturbations in the density field are inherently
triaxial.
After turnaround, each axis may evolve separately  until the final axis collapses and the
object virialises.
 \citet{Sheth:1999su} and \citet{Sheth:2001dp} (hereafter S-T) modified the P-S formalism, 
replacing the spherical collapse model  with ellipsoidal collapse, in which the
surrounding shear field as well as the initial overdensity
determines the collapse time of an object.
\citet{Sheth:1999su} found a universal mass function for any CDM model.
\citet{Jenkins:2000bv} found a universal empirical fit to the form of the mass function measured from a suite of cosmological 
simulations.
The Jenkins et al.  mass function 
can accurately predict halo abundances over a range of cosmologies
and redshifts (see also \citealt{Warren:2005ey,Reed:2006rw} and \citealt{Crocce:2009mg}).

We use a friends-of-friends (FOF) halo finder, with a constant linking length of $b=0.2$, to identify haloes in all cosmologies. 
In Fig. \ref{J01} we plot groups containing 20
particles or more to ensure that the systematic uncertainties in the mass function are at
or below the 10\% level; tests show that 90\% or more of such haloes are gravitationally bound \citep{Springel:2005nw}.
The first row in Fig. \ref{J01} shows the mass function for SUGRA and $\Lambda$CDM at $z = 0$, 1 and 2.  
 The filled red squares represent the mass function from Stage III of the simulations 
where a consistent linear theory and cosmological parameters were used for the SUGRA model.
The mass function for $\Lambda$CDM (open black circles) and
the SUGRA  model are plotted together with the Jenkins et al. mass function shown in black (red) for $\Lambda$CDM (SUGRA).
The S-T mass function is shown in the top left panel in the first row of this figure (blue dashed line) for comparison.
The abundances in both $\Lambda$CDM and SUGRA agree with each other at redshift 0 and with the Jenkins et al. and S-T models, although
the fitting formulae seem to slightly under-predict the number of haloes at the high mass end
($M > 10^{15} h^{-1} \, M_{\sun}$).
In the first row of Fig. \ref{J01}, the number of haloes in the two models start to differ at $z=1$,  and at $z=2$
there is a large difference in the mass functions.
The linear growth factor for the SUGRA model together with the best fit cosmological
parameters from Table A3 have been used to obtain the
Jenkins et al. fit  at the earlier redshifts.
The Jenkins et al. fit describes the data slightly better at the high mass end
at higher redshifts than the S-T prescription. This is as expected  as the Jenkins et al. fit was
explicitly tested  at the high mass end of the mass function.
Each model shows only small ($<20$\%) differences
between the measured value and the Jenkins et al. fitting formula for $M < 10^{15} h^{-1} M_{ \sun}$ at $z=0$.
Underneath each panel in the first row in Fig. \ref{J01}, we plot the ratio between the measured mass function  for $\Lambda$CDM and the SUGRA model in Stage III, 
and the Jenkins at al. mass function for $\Lambda$CDM.

The second row of Fig. \ref{J01} repeats this comparison for the AS model.
In this row the mass function for $\Lambda$CDM (open black circles) and the AS model from Stage III
 (green squares) of the simulations at $z=0$, 1 and 2 are plotted. The Jenkins et al. 
mass function for $\Lambda$CDM (black line) and the AS model for Stage III (green line) are also plotted.
The AS model has  a greater abundance of halos than $\Lambda$CDM at $z=2$. For the Stage III simulation,
 the AS model has $\Omega_{\rm m}
h^2 = 0.086$ giving rise to a change in the spectral shape of the linear theory power spectrum from $\Lambda$CDM linear theory
($\Omega_{\rm m}
h^2 = 0.133$). 
As a result there are fewer low mass halos and a similar number of high mass haloes at $z=0$ compared to $\Lambda$CDM.
This change accounts for the
 decrease in the mass function for $M<10^{15} h^{-1} M_{\sun}$ seen at $z=0$ in the AS model (green squares).
At $z=0$, there are only small ($<20$\%) differences
between the measured value and the Jenkins et al. fitting formula for $M < 10^{15} h^{-1} M_{\sun}$ for $\Lambda$CDM and the AS model from Stage III.
The ratio between the Jenkins et al. 
mass function  for $\Lambda$CDM and the measured mass function for $\Lambda$CDM and the AS model 
from Stage III is plotted underneath each panel in the second row in Fig. \ref{J01}.  

Only the SUGRA and AS models are plotted in Fig. \ref{J01} but similar differences in halo abundances are seen in the  INV
models
compared to $\Lambda$CDM, whilst only negligible differences with $\Lambda$CDM were found in the mass functions
of
2EXP and CNR.
\citet{Grossi:2008xh} found similar results for the mass function 
over the range $10^{11}$ - $10^{14}h^{-1}M_{\sun}$ in an \lq early dark energy\rq \, model,
using much smaller volume simulations than ours. They found a higher number density
of haloes corresponding to groups and clusters in non-standard dark energy models at high redshifts compared to $\Lambda$CDM, while at $z=0$ the models all
agreed with one another. We find similar results 
although using the cosmological parameters from Table A3 for each quintessence model can give different abundances at $z=0$ in those models 
compared to $\Lambda$CDM because although $\sigma_8$ is the same the shape of the linear theory can be different.
Also, we have been able to  probe a higher mass range for the dark matter haloes. The high mass end  of the
mass function is very  sensitive to changes in the current value of the linear growth factor in the different cosmologies.

\begin{figure*}
\center
{\epsfxsize=15.truecm
\epsfbox[101 370 523 613]{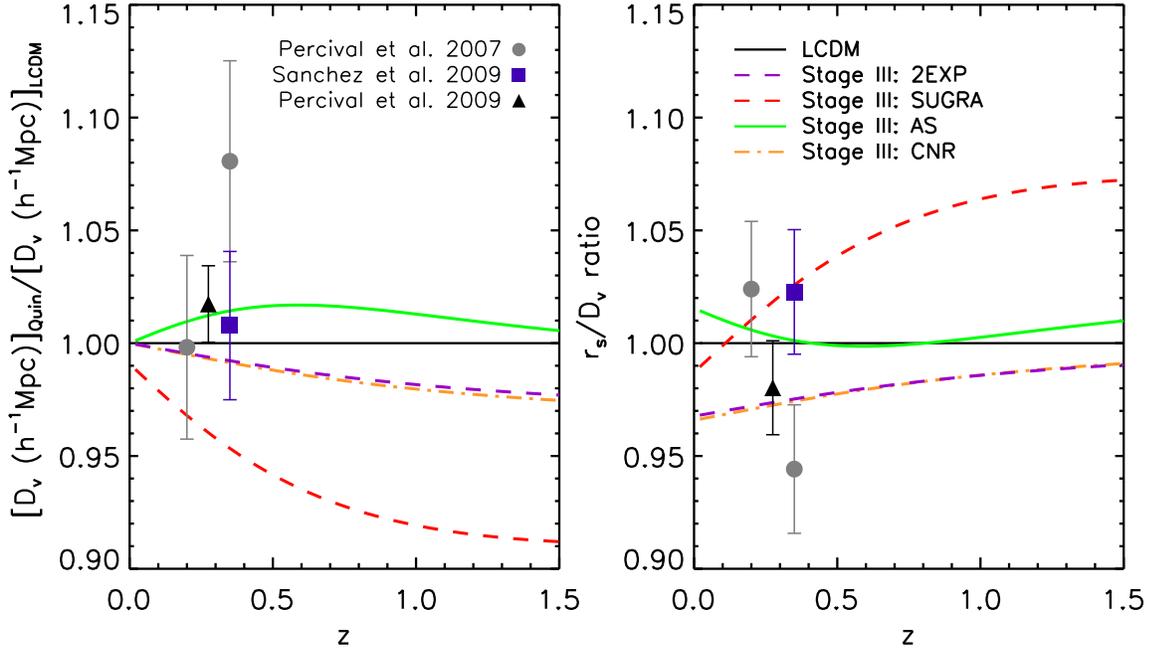}}
 \caption{ The ratio of the distance measure $D_v(z)$ (left panel) and the ratio of $r_s(z_d)/D_v$ (right panel) for four quintessence models compared to $\Lambda$CDM
 as indicated by the key in the right hand panel.
The grey circles are estimate points from \citet{Percival:2007yw} at $z=0.2$ and $z=0.35$ measured using the observed scale of BAO calculated from the SDSS and 2dFGRS
main galaxy samples. \citet{Sanchez:2009jq} combined CMB data with information on the shape of the redshift space correlation function using a larger
LRG
dataset and found
$D_v(z=0.35) = 1300 \pm 31$ Mpc and $r_s(z_d)/D_v = 0.1185 \pm  0.0032$ at $z=0.35$ (blue squares).
The data points from \citet{Percival:2009xn} for $D_v$ and $r_s(z_d)/D_v$ at $z=0.275$ using WMAP 5 year data + SDSS DR7 are plotted as black triangles.
}\label{RSDV}
\end{figure*}

In Fig. \ref{universal} we  plot the fraction of the total mass in haloes of mass $M$ rather than simply the abundance as shown in Fig. \ref{J01}.
We compare the Jenkins et al. analytic fit to our simulated halo mass functions in the 
SUGRA and AS models and in $\Lambda$CDM at $z=0$ and 1 in Fig. \ref{universal}. In this plot the  quantity $\mbox{ln}  \sigma^{-1} (M, z)$ is used 
 as the mass variable
instead of $M$, where $\sigma^2 (M, z)$ is the variance of the linear
density field at $z=0$.
This variance can be expressed as
\begin{equation}
\sigma^2 (M,z) = \frac{D^2(z)}{2\,\pi^2}\, \int_0^{\infty} k^2 P(k) W^2(k;M) {\rm d}k \, ,
\end{equation}
where $W(k;M)$ is a top hat window function enclosing a mass $M$, $D(z)$ is the
linear growth factor of perturbations at redshift $z$ and $P(k)$ is the power
spectrum of the linear density field.
Plotting different masses at different redshifts in this way takes out the redshift dependence in the power spectrum. Note a large value of $ \mbox{ln}  \sigma^{-1} (M, z)$ 
corresponds to a rare halo.
 Using this variable, Jenkins et al.
 found that the mass function
at different epochs has a universal form, for a fixed power spectrum shape. Note that in our case, the Stage III simulations have somewhat different 
power spectra, which account for the bulk of the dispersion between the simulation results at the rare object end of Fig. \ref{universal}; in Stage I, 
the simulation results agree with the Jenkins et al. universal form to within $ 25\%$ at $\mbox{ln}  \sigma^{-1} = 1.0$.
As shown in Fig. \ref{universal}, we find the Jenkins et al. fitting formula is accurate to $\sim 20\%$ at $z=0$ for all the models 
in the range $M<10^{15}h^{-1}M_{\sun}$. At higher redshifts the measured mass function for the SUGRA model and $\Lambda$CDM differ from the 
Jenkins et al. mass function by $\sim 30\%$ over the same mass range while for the AS model the difference is $\sim 50\%$ at $z=1$.
In previous work, \citet{Linder:2003dr} also found that the predicted mass function
for a SUGRA-QCDM simulation, which would be the equivalent of our Stage I simulations, was well fit (within 20\%) by the 
Jenkins et al. formula.

\subsection{The appearance of baryonic acoustic oscillations in quintessence models\label{bao}}

In this section we examine the baryonic acoustic oscillation signal in the matter power spectrum for the AS, SUGRA and CNR models. 
\citet{Angulo:2007fw} presented a detailed set of predictions for the appearance of the BAO signal in the $\Lambda$CDM model,
covering the impact of nonlinear growth, peculiar velocities and scale dependent redshift space distortions and galaxy bias. Here we 
focus on the first of these effects and show power spectra in real space for the dark matter.
We do not consider the INV1 model as it is not consistent with observational constraints (Appendix A), or the INV2 or 2EXP models as they are
 indistinguishable from $\Lambda$CDM, and hence were not simulated again in Stage III (Section \ref{wmap}). 

In Stage I of our simulations (Section \ref{psI}), we would expect the linear theory comoving BAO for the quintessence models 
to be identical to $\Lambda$CDM as the same linear theory power was used for all models. 
In Stage II (Section \ref{psII}), some of the quintessence models have large amounts of dark energy at early times which will alter the sound horizon 
in these models compared to $\Lambda$CDM (see Table A3), and as a result we would expect to see a corresponding 
shift in the BAO peak positions.
The best fit cosmological parameters found in Stage III were derived using CMB, BAO and SN distance measurements (see Appendix \ref{A}).
 Stage III of our simulations (Section \ref{wmap}) uses these parameters and  we would expect models with the same BAO 
distance measures to have the same peak pattern in the matter power spectrum as $\Lambda$CDM.

The baryonic acoustic oscillations are approximately a standard ruler and depend on the sound horizon, $r_s$, given in Eq. \ref{soundhorizon} \citep{Sanchez:2008iw}.
 The apparent 
size of the BAO scale depends on the distance to the redshift of observation and on the ratio $r_s/D_v$, where $D_v$ is an effective distance measure 
which is a combination of $D_A$ and $H$, given in Eq. \ref{DV}.
In most quintessence models, $r_s$ remains unchanged unless there is appreciable dark energy at last scattering. Models which have the same 
ratio of $r_s/D_v$ are impossible to distinguish using BAO.

\begin{figure*}
\center
{\epsfxsize=17.truecm
\epsfbox[84 366 536 612]{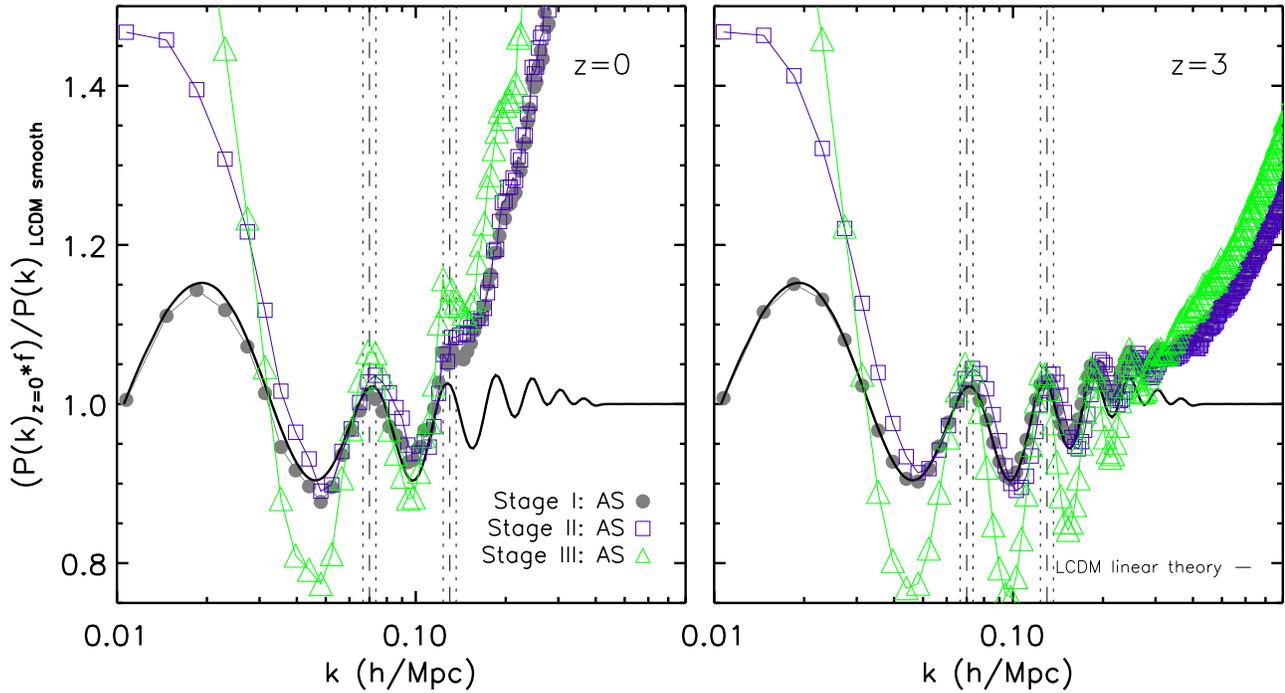}}
 \caption{The real space power spectrum for the AS model on large scales at $z=0$ (left) and
$z=3$ (right). All power spectra have been divided by a smoothed
linear \lq no-wiggle\rq \, theory $P(k)$ for $\Lambda$CDM. The factor, $f$, removes the scatter of the power measured in the simulation
around the expected linear theory power.
 Stage I in our simulation is represented by  grey circles, Stage II is represented by open
blue squares and Stage III results are shown as
 green triangles. The black solid line represents the linear theory power spectrum in $\Lambda$CDM divided by the smooth reference spectrum.
The vertical dashed (dotted) lines show the position of the first two  acoustic
peaks (positions $\pm 5\%$) for a $\Lambda$CDM cosmology.
}\label{BAOI}
\end{figure*}
\begin{figure*}
\center
{\epsfxsize=17.truecm
\epsfbox[84 366 536 612]{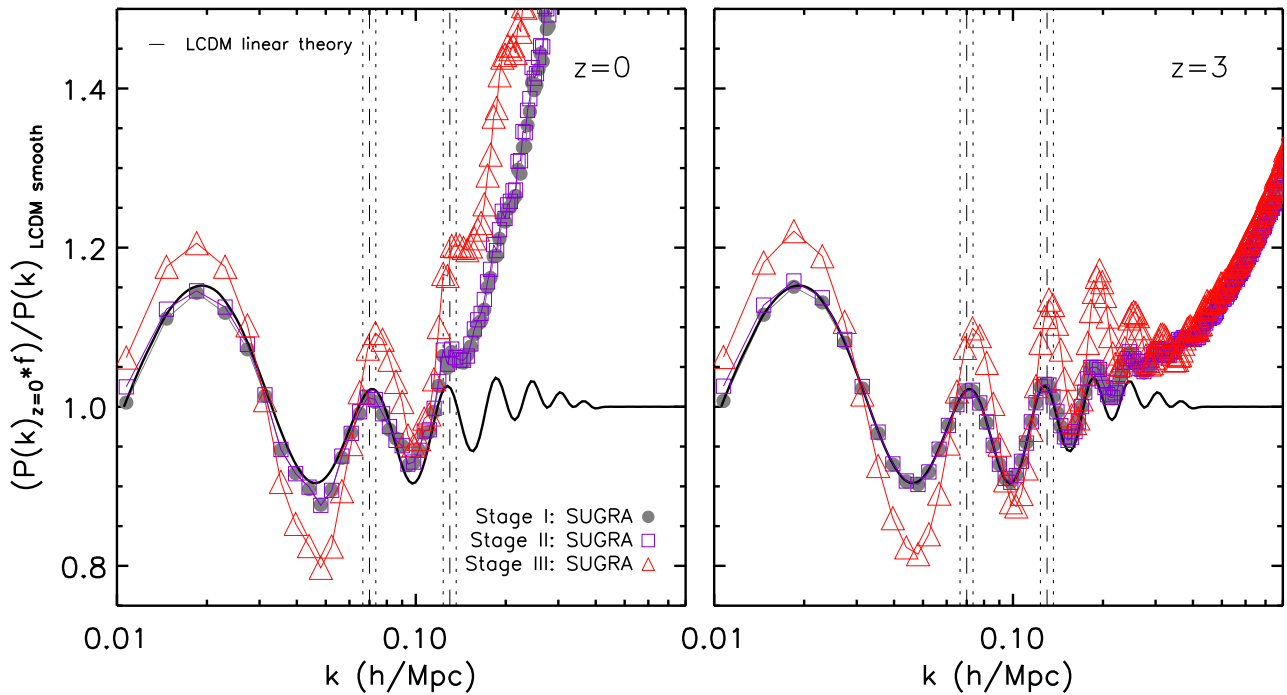}}
 \caption{ The real space power spectrum for the SUGRA model on large scales
at $z=0$ and $z=3$. All power spectra have been divided by a smoothed
linear theory $P(k)$ for $\Lambda$CDM.
 Stage I in our simulation is represented by  grey circles, Stage II is represented by open purple squares
and Stage III results are shown as
 red triangles. The black solid line represent the linear theory power spectrum in $\Lambda$CDM divided by the smooth reference spectrum.
The vertical dashed (dotted) lines show the position of the first
two acoustic peaks (positions $\pm 5\%$)
for a $\Lambda$CDM cosmology.
}\label{BAOIII}
\end{figure*}

To calculate the power spectrum for a galaxy redshift survey, the measured angular and radial separations of galaxies pairs are converted to
co-moving separations and scales. This conversion is dependent on the cosmological model assumed in the analysis.
These changes can be combined into the single effective measure, $D_v$.
Once the power spectrum is calculated in one model we can simply re-scale $P(k)$ using $D_v$ to obtain the
power spectrum and BAO peak positions in another cosmological model (see \citealt{Sanchez:2009jq}).
In the left panel of Fig. \ref{RSDV}, we plot the ratio of $D_v$ in four quintessence models compared to $\Lambda$CDM up to $z=1.5$.
\citet{Percival:2007yw} found $D_v = 564 \pm 23 h^{-1}$Mpc at $z=0.2$ and $D_v = 1019 \pm 42 h^{-1}$Mpc at $z=0.35$ using
the observed scale of BAO measured from the SDSS DR5 galaxy sample and 2dFGRS.
These data points are plotted as grey circles in Fig. \ref{RSDV}. 
Note that at face value none of the models we consider are consistent with the \citet{Percival:2007yw} point at $z=0.35$.
These authors report a 2.4$\sigma$ discrepancy between their results using BAO and the constraints available at the time from supernovae. 
The blue square plotted in the left panel in  Fig. \ref{RSDV} is
the constraint  $D_v = 1300 \pm 31$ Mpc at $z=0.35$ found by \citet{Sanchez:2009jq}. This constraint was found using a much larger
LRG
dataset and improved modelling of the correlation function on large scales. The constraint found by \citet{Sanchez:2009jq}
using CMB  and BAO data is fully consistent with CMB and SN results.
The results from \citet{Percival:2009xn} for $D_v$ and $r_s(z_d)/D_v$ at $z=0.275$ using WMAP 5 year data together with the SDSS data release 7 galaxy sample are also plotted
(black triangles). The \citet{Percival:2009xn} results are in much better agreement with those of \citet{Sanchez:2009jq}.

Over the range of redshifts plotted in Fig. \ref{RSDV} the distance measure, $D_v$, in the AS, 2EXP and CNR models differ from $\Lambda$CDM by at most 2\% and is $<1\%$ in these
models for $z<0.2$. Re-scaling the power spectrum for these dark energy cosmologies
 would result in a small shift $\sim 1\%$ in the position of the peaks at low redshifts.
The value of $D_v$ in the SUGRA model differs from $\Lambda$CDM by at most 9\% up to $z=1.5$.
The right panel in Fig. \ref{RSDV} shows the ratio of $r_s(z_d)/D_v$ in the quintessence models compared to $\Lambda$CDM, where $r_s$ is the
co-moving sound horizon scale
at the drag redshift, $z_d$, which we discuss in Appendix \ref{A}.
The value of $r_s(z_d)/D_v$ can be constrained using the position of the BAO in the power spectrum. In the right panel of Fig. \ref{RSDV} the grey symbols are the results from
\citet{Percival:2007yw} at $z=0.2$ and $z=0.35$. From this plot it is clear that the SUGRA and AS model are within the 1$\sigma$ limits at $z=0.2$.
The 2EXP and CNR model lie just outside the 1$\sigma$ errors at $z=0.35$. Note the value of $r_s(z_d)/D_v$ for $\Lambda$CDM at $z=0.35$ also lie
outside the  1$\sigma$ errors, see \citet{Percival:2009xn} for more detail. The blue square plotted in the right panel in 
Fig. \ref{RSDV} is $r_s(z_d)/D_v= 0.1185 \pm  0.0032$ at $z=0.35$ and was obtained using  
information on the redshift space correlation function together with CMB data \citep{Sanchez:2009jq}.

In Fig. \ref{BAOI} and \ref{BAOIII}
 we plot the $z=0$ and $z=3$ power spectra in the AS and SUGRA models divided 
by a linear theory $\Lambda$CDM reference spectrum  which has been smoothed using the coarse rebinning method proposed by
 \citet{Percival:2007yw} and refined by \citet{Angulo:2007fw}. After dividing 
by this  smoothed  power spectrum, the acoustic peaks are more visible in the quasi-linear regime.
In Figs. \ref{BAOI} and \ref{BAOIII}, the 
measured power in each bin has been multiplied by a factor, $f$, to  
remove
the scatter due to the small number of large scale modes in the simulation \citep{1994MNRAS.270..183B,Springel:2005nw}. 
This factor, $f = P(k)_{\tiny \mbox{linear}}/P(k)_{\tiny \mbox{N-body}}$, 
is the ratio of the expected linear theory power and the measured power in each bin at $z=5$,
at which time  the power on these scales is still expected to be
linear.
Multiplying by this correction factor allows us to see 
the onset of nonlinear growth around $k \sim 0.15 h$Mpc$^{-1}$ more clearly.

In Fig. \ref{BAOI} (\ref{BAOIII}) we plot the AS (SUGRA) power spectrum  as grey circles from Stage I, blue (purple) squares from 
Stage II and green (red) triangles from Stage III. The black line represents
 the linear theory power in $\Lambda$CDM divided by the smooth reference spectrum.
In both plots and for all power spectra, the same reference spectrum is used. The reference is a simple \lq wiggle-free\rq \, CDM 
spectrum, with a form controlled by the shape parameter $\Gamma = \Omega_{\rm m} h$ \citep{1986ApJ...304...15B}.
 The difference between the 
AS and $\Lambda$CDM linear theory, as shown in Fig. \ref{PlotPk}, results in an increase 
in large scale power on scales $k<0.04 h$Mpc$^{-1}$.
The vertical dashed (dotted) lines show the first two positions of the acoustic peaks 
(positions $\pm 5\%$) for a $\Lambda$CDM cosmology.

As shown in Fig. \ref{BAOI}, we find that the position of the first acoustic peak in the AS  model from Stage I is the same as 
 in $\Lambda$CDM.
 The position of the first peak for the AS model, measured in 
Stage II of our simulations (blue squares),
  is slightly shifted ($\sim 4\%$) to 
smaller scales compared to $\Lambda$CDM as the sound horizon is altered in the AS model. 
In Stage III, when the best fit  cosmological parameters for the AS model are used, the sound 
horizon in the AS model and  $\Lambda$CDM are very similar at $z \sim 1090$ and there is a very small ($<1$\%)
 shift in the position of the first peak (green triangles). As 
there is less nonlinear growth at $z=3$ the 
 higher order peaks are more visible in the right-hand plot in Fig. \ref{BAOI}.

In Fig. \ref{BAOIII}, the SUGRA power spectrum from Stage I, II and III are plotted. The SUGRA $P(k)$ from Stages I and II have identical 
peak positions to $\Lambda$CDM as the sound horizon is the same as in  $\Lambda$CDM in these cases.
There is a shift ($\sim 5\%$) in the position of the
  first peak in the SUGRA model using the $P(k)$ measured in Stage III.
Note the units on the x axis are $h/$Mpc and from Table A3, $h=0.67$ for the 
Stage III SUGRA model  compared to $h=0.715$ for $\Lambda$CDM.
On small scales 
the BAO signature is damped due to more nonlinear  structure formation at $z=0$ compared to $z=3$ as shown in Fig \ref{BAOIII}. 
We find a large increase in the power in the region of the
 second peak, $k \sim 0.15 h$Mpc$^{-1}$ in both the AS and SUGRA models, measured in Stage III, compared to $\Lambda$CDM.
For brevity we have not included the plots of the power spectra for the CNR model showing the baryonic acoustic oscillations. We find identical peak positions 
in $\Lambda$CDM and this model in all stages  at $z=0$.

The AS and SUGRA model are very different to $\Lambda$CDM at late times and as result they affect the growth of structure at $z>0$ as seen in Sections \ref{wmap} 
and \ref{mf}. We have found that models like this do not necessarily have  different BAO peak positions to $\Lambda$CDM in the matter power spectrum.
These results  suggest that 
distinguishing a quintessence model, like the AS  model used in this paper, using measurements of the BAO peak positions in future 
galaxy surverys, will be extremely difficult. The BAO peak positions for the  CNR model will be shifted by at most 2\% in the range $z<1.5$
compared to $\Lambda$CDM after re-scaling the power spectra by $D_v$. In conclusion it is possible to have  
quintessence cosmologies with  higher levels of dark energy at early times than in $\Lambda$CDM 
and still measure the same peak positions for the BAO in the matter power spectrum.

\section{Conclusions and Summary}
 
Observing the dynamics of dark energy is the central goal of future galaxy surveys and would distinguish a cosmological constant from a dynamical quintessence model.
Using  a broad range of quintessence models, with either a slowly or rapidy varying equation of state,
we have analysed the influence of dynamical dark energy on structure formation using N-body simulations.

We have considered a range of quintessence
models that can be classified as either
\lq tracking\rq \, models, for example the SUGRA and INV models, or \lq scaling\rq \, solutions, such
as the AS, CNR or 2EXP models, depending on the
evolution of their equation of state (see Table 1 and Section 2.1).
The models feature both rapidly and slowly
varying equations of state and the majority of the
models could be classified as \lq early dark energy\rq \, models as they have a non-negligible amount of
dark energy at early times.

\begin{table*}
\caption{ The key features in the evolution of the quintessence models simulated. $\Delta D (z=5) $ is the ratio of the linear growth factor for each quintessence
model compared to $\Lambda$CDM at $z=5$. A late time transition in the equation of state is defined as occuring at $z<2$. The AS, CNR, 2EXP and SUGRA models can be considered as \lq early 
dark energy \rq \, models as they have non-negligible amounts of dark energy present at early times.
}
\begin{tabular}{|c|r|r|r|r|}
\hline
Model & transition type& transition redshift &  $\Omega_{\tiny \mbox{DE}}(z=300)$  &  $\Delta D (z=5) $    \\
\hline \hline
INV1 & gradual & $\sim 4.5$  & $\sim 0.009$ & $\sim 50\%$   \\
INV2 & gradual & $\sim 5$  & negligible & $\sim 10\%$   \\
SUGRA & rapid & $\sim 9$  & $\sim 0.01$ & $\sim 20\%$   \\
2EXP & rapid & $\sim 4$  & $\sim 0.015$ & $0\%$   \\
CNR & rapid & $\sim 5.5$  & $\sim 0.03$ & $0\%$   \\
AS & rapid & $\sim 1$  & $\sim 0.11$ & $20\%$   \\
\hline
\end{tabular}
\end{table*}

In order to accurately mimic the dynamics of the original quintessence 
models at high and low redshift, it is necessary to use a general 
prescription for the dark energy equation of state 
which has more parameters than the ubiquitous 2 variable equation. Parametrisations for $w$ which use
 2 variables are unable to faithfully 
represent dynamical dark energy models  over a wide range of redshifts  and can lead to biases when used to constrain parameters \citep{Bassett:2004wz}. 
Our task has been made easier by the availability of parametrizations which accurately describe the dynamics of the different quintessence models
\citep{Corasaniti:2002vg, Linder:2005ne}.
This allows us to modify the Friedmann equation in the simulation, using the equation of state as a function of redshift. We use the 
parametrization of \citet{Corasaniti:2002vg}.
In its full six parameter form, this framework can describe the quintessence model back to the epoch of nucleosynthesis. Four parameters are 
sufficient to describe the behaviour of the quintessence field over the redshift interval followed by the simulations.
With this description of the equation of state, our simulations are able to accurately describe the impact of the quintessence model on the expansion 
rate of the Universe, from the starting redshift to the present day. This would not be the case with a 2 parameter model for the equation of state.

In this paper we have taken into account three levels of modification from a
$\Lambda$CDM cosmology which are necessary if we wish to faithfully incorporate 
the effects of quintessence dark energy into a N-body simulation.
The first stage is to replace the cosmological constant with the quintessence model in the Friedmann equation. A quintessence model with a different 
equation of state from $w=-1$ will lead to a universe with a different expansion history. This in turn alters the rate at which perturbations can collapse under gravity.
The second stage is to allow the change in the expansion history and perturbations in the quintessence field to have an impact on the form of the 
linear theory power spectrum.
 The  shape of the power spectrum can differ significantly from 
$\Lambda$CDM on large scales if there is a non-negligible 
amount of dark energy present at early times.
This alters the shape of the turn-over in the power spectrum compared to $\Lambda$CDM.
Thirdly, as the quintessence model should be consistent with observational constraints, 
the cosmological parameters used for the dark energy model could be different from the 
best fit $\Lambda$CDM 
parameters.
In the three stages of simulations we look at the 
effect each of the above modifications has on the nonlinear growth of structure. 
Deconstructing the simulations into three stages allows us to isolate specific features in the 
quintessence models  which play a key 
role in the growth of dark matter perturbations.

In the first stage of comparison, in which all that is changed is the expansion history of the universe, we found that some of the quintessence models showed
enhanced structure formation at $z>0$ compared to $\Lambda$CDM.
The  INV1, INV2, SUGRA and AS models  have slower 
growth rates than $\Lambda$CDM. Hence, when normalising to the same $\sigma_8$ today, 
structures must form at earlier times in these models to 
overcome the lack of  growth at late times.
Models such as the 2EXP and 
CNR model have the same recent 
growth rate as $\Lambda$CDM and showed no difference in the growth of structure.
The difference in linear and nonlinear growth can largely be explained by the difference in the growth factor at different epochs in the models.
At the same growth factor, the power in the models only diverges at the 15\% level well into the nonlinear regime.

In the second stage, 
a self-consistent linear theory $P(k)$ was used for each quintessence model 
to generate the 
initial conditions in the simulations. The amount of dark energy present at early times will determine
the  impact on the linear theory dark matter power spectrum and the magnitude of deviation from the 
$\Lambda$CDM spectrum. 
High levels of dark energy at early times suppress
the growth of the dark matter on scales inside the horizon, 
resulting in a broader turn-over in the power spectrum.
We found that models with the highest levels of dark energy at the last scattering surface, such as the AS and CNR models, 
have linear theory $P(k)$ which differ the most from $\Lambda$CDM. 
The results of the N-body simulations of the AS and the SUGRA model show a very small increase in nonlinear 
growth compared to the results in Stage I. The increase in the linear theory power is on 
very large scales and does not change the small scale growth significantly.

In our final stage of simulating the effects of 
quintessence, we found the best fitting 
cosmological parameters for each model, $\Omega_{\rm m}
h^2$, $\Omega_{\rm b}
h^2$ and $H_0$, consistent with current CMB, SN and BAO meaurements. 
For quintessence dark energy models, it is important to consider the changes in more than just one cosmological parameter when fitting to the observational data.  
For example, for a given dark energy equation of state,
the values of $\Omega_{\rm m}
h^2$ and $H_0$ may change in such a way to compensate one another and give similar growth rates and expansion histories to
$\Lambda$CDM. These compensating effects will be missed if, for example, only 
$\Omega_{\rm m}$ is changed for the dark energy model as in recent work
\citep{2009arXiv0903.5490A}.
Models with cosmological parameters which fit the data but were significantly 
different from $\Lambda$CDM were simulated again (Section \ref{wmap}).

We will now summarise and discuss the main results for each model. The key features of each of the quintessence models are presented in Table 2.
The INV1 model was unable to fit the data with a reasonable $\chi^2/\nu$ (Table A3). 
This toy model had the largest growth factor ratio to $\Lambda$CDM at $z=5$ and 
as a result showed the most enhanced growth in Stage I of our simulations.
The 
 linear growth factor for the INV2 model is very different to $\Lambda$CDM at early times and
 gives rise to enhanced growth at $z>0$ as seen in Section \ref{psI}. 
This model has negligible dark energy at early times and so the 
spectral shape is not altered in Stage II. 
In the 2EXP model the rapid transition to $w=-1$ in the equation of state early on leaves little impact on the growth of dark matter and as a result the
power spectra and mass function are indistinguishable from $\Lambda$CDM.
As both the INV2 and 2EXP models already
agree with cosmological measurements with very similar values 
for $\Omega_{\rm m}
h^2$, $\Omega_{\rm b}
h^2$ and $H_0$ to $\Lambda$CDM, we did not run these simulations again. 

The SUGRA model has   
enhanced linear and nonlinear growth and halo abundances compared to 
$\Lambda$CDM at $z>0$ and an altered linear theory power spectrum shape. 
The mass function
results for all stages of our simulations for the SUGRA model show enhanced halo abundances at $z>0$.
Analysing the SUGRA power spectra,
 from a Stage III simulation which used the best fit parameters for this model, reveals a $\sim 5\%$ shift in the position of the first BAO 
peak. We find the distance measure $D_v$ for the SUGRA model 
differs by up to 9\% compared to $\Lambda$CDM  over the range $0<z<1.5$. Re-scaling the power measured for the SUGRA model by the difference in $D_v$ 
would result in an even larger shift in the position of the 
BAO peaks.

The CNR model has high levels of dark energy early on
which alters the spectral shape on such large scales that the nonlinear growth of structure is
only slightly less than $\Lambda$CDM at $z<5$.
This model has a halo mass abundance at $z<5$ and BAO peak positions at $z=0$ which are the same as in $\Lambda$CDM. 
For $z<0.5$ the distance measure, $D_v$, for the CNR model 
differs from  $\Lambda$CDM by $\sim 1\%$, as result there would be a corresponding small shift in the BAO peak positions. 
The rapid early transition at $z=5.5$ in the equation of state to $w_0=-1$ in this model seems to remove any signal of the large amounts of 
dark energy at early times that might be present in the growth of dark matter perturbations.

The AS model has the highest levels of dark energy at early times, and so 
its linear theory spectrum is altered the most. 
This results in a large increase in large scale power, when we normalise the power spectrum to $\sigma_8 =0.8$ today. 
The results from Stage III using the best fit parameters show both enhanced linear and nonlinear growth at $z<5$. The linear theory $P(k)$ is altered on scales $k\sim 0.1 h$Mpc$^{-1}$ which drives an increase in nonlinear growth on small scales compared to $\Lambda$CDM. The mass function 
results in Stage III for this model show enhanced halo abundances at $z>0$. 
We find that using the best fit cosmological parameters for the  AS model produces a BAO profile
 with peak positions similar to those in $\Lambda$CDM. At low redshifts there is  $\sim 1\%$ shift in the first peak compared 
to $\Lambda$CDM  after re-scaling the power with the difference in the distance measure $D_v$ between the two cosmologies.

These results from Stage III of our N-body simulations 
show that dynamical dark energy models in which the dark energy equation of state makes a late $(z<2)$ rapid transition to $w_0=-1$ show enhanced linear and nonlinear growth compared to  $\Lambda$CDM at $z>0$ and have a greater abundance of dark matter haloes compared to  $\Lambda$CDM for $z>0$. 
We found that dynamical dark energy 
models can be significantly different from $\Lambda$CDM at late times and still produce similar BAO peak positions in the matter power spectrum.
Models which have a rapid early transition in their dark energy equation of state and mimic $\Lambda$CDM after the transition, show the same linear and nonlinear growth and 
halo abundance as $\Lambda$CDM for all redshifts. We have found that these models can give rise to BAO peak positions
in the matter power spectrum which are the same as those
 in a $\Lambda$CDM cosmology. This is true despite these models having non-negligible amounts of dark energy present at early times.

Overall, our analysis shows that the prospects of detecting dynamical dark energy, which features a late time transition, using the halo mass function at $z>0$ are good, 
provided a good proxy can be found for mass. Parameter degeneracies allow some quintessence models 
to have identical BAO peak positions to $\Lambda$CDM and so these measurements alone will not be able 
to rule out some quintessence models.
Although including the dark energy
perturbations has been found to increase 
these degeneracies \citep{Weller:2003hw}, incorporating them into the N-body code  would clearly be 
the next step towards simulating quintessential dark matter 
with a  full physical model. 
Although in many quintessence models the dark energy 
clusters on very large scales today ($k<0.02h$Mpc$^{-1}$) \citep{Weller:2003hw}
 and the perturbations are generally small ($\delta_{\tiny \mbox{DE}} \sim 10^{-1}$),  these 
perturbations may nevertheless have some impact on the dark matter structure  
in a full 
N-body simulation of the nonlinear growth.

\section*{Acknowledgments}

EJ acknowledges receipt of a fellowship funded by the European
 Commission's Framework Programme 6, through the Marie Curie Early Stage
Training project MEST-CT-2005-021074.
This work was supported in
part by grants from the Science and Technology Facilities
Council held by the Extragalactic Cosmology Research Group and 
the Institute for Particle Physics Phenomenology at Durham University.
We acknowledge helpful conversations with Simon D. M. White, Ariel G. S\'{a}nchez, Shaun Cole and Lydia Heck for support running the simulations.

\bibliographystyle{mn2e}
\bibliography{mybibliography}

\appendix
\section{Observational distance priors \label{A}}

\begin{table*}
\label{wmapfixed}
\caption{Distance priors based on WMAP observations \citep{Komatsu:2008hk} for each quintessence model using $\Omega_{ \rm m}
h^2$, $\Omega_{ \rm b}
h^2$ and $H_0$  parameters from
\citet{Sanchez:2009jq}. These parameters were derived assuming a $\Lambda$CDM cosmology.
$l_A(z_*)$ is the acoustic scale at the epoch of decoupling, $z_*$ and $R(z_*)$ is the shift parameter.
$ \chi^2_{\tiny \mbox{total}}=\chi^2_{\tiny \mbox{WMAP+SN+BAO}}$ and $\nu$ is the number of degrees of freedom.}
\begin{tabular}{|c|r|r|r|r|}
\hline \hline
 & $z*$  &   $l_A(z*)$ & $R(z*)$  &$\chi^2_{\tiny \mbox{total}}/\nu$ \\
\hline
WMAP 5-yr ML & 1090.51 $\pm 0.95$ &  302.10 $\pm$ 0.86 & 1.710 $\pm$ 0.019 & 0\\
\hline
INV1 &-  & 261.05 & 1.49 &15.34\\
INV2 &-   & 294.34 & 1.67 &1.81 \\
SUGRA &-  & 284.03 & 1.62 & 3.88\\
2EXP &-  & 303.85 & 1.74  & 1.09\\
AS & -  & 289.69& 1.74 & 2.04\\
CNR &-   & 306.71 & 1.79 &1.37\\
\hline
\end{tabular}
\end{table*}

\begin{table*}
\label{baofixed}
\caption{BAO distance measurements \citep{Percival:2007yw} for each quintessence model using $\Omega_{\rm m}
 h^2$, $\Omega_{\rm b}
h^2$ and $H_0$  parameters from
\citet{Sanchez:2009jq}. These parameters were derived assuming a $\Lambda$CDM cosmology. A fitting formula proposed by \citet{Eisenstein:1997ik} was used for the drag redshift $z_{\tiny \mbox{drag}}$.}
\begin{tabular}{|c|r|r|r|r|}
\hline \hline
 & $z_{\tiny \mbox{drag}}$  & $r_s(z_{\tiny \mbox{drag}})$ & $r_s(z_{\tiny \mbox{drag}})/D_v(z=0.2)$  & $r_s(z_{\tiny \mbox{drag}})/D_v(z=0.2)$    \\
\hline
WMAP 5-yr & 1020.5 $\pm$ 1.6 & 153.3 $\pm$ 2.0 Mpc & - &- \\
\citet{Percival:2007yw} & - & 154.758 Mpc & 0.198 $\pm$ 0.0058 & 0.1094 $\pm$ 0.0033 \\
\hline
INV1 & - & 152.5 Mpc & 0.208 &  0.130 \\
INV2 & -& 152.7 Mpc & 0.198 &  0.121 \\
SUGRA & -& 152.5 Mpc & 0.198 & 0.121 \\
2EXP & -& 152.0 Mpc &  0.192 & 0.115 \\
AS &-& 143.9 Mpc & 0.183 & 0.111 \\
CNR &-& 150.7 Mpc & 0.191 & 0.114 \\
\hline
\end{tabular}
\end{table*}

In this section we outline the method used to find the best fit cosmological parameters for each of the quintessence models using CMB, BAO and SN data. 
The method suggested in
\citet{Komatsu:2008hk} employs three distance priors from measurements of the CMB together
with the \lq UNION\rq\,  supernova samples \citep{Kowalski:2008ez} and the baryonic acoustic oscillations (BAO)
in the distribution of galaxies \citep{Percival:2007yw}
to explore the best fit parameters for the dynamical dark energy models.
In Section \ref{psI} and \ref{psII}, all of the quintessence simulations were run using the best fit cosmological parameters
assuming a $\Lambda$CDM model. While this is useful for isolating the effect of the different expansion histories
on the growth of structure, this does not yield  quintessence models which would 
automatically satisfy
the constraints on distance measurements.
Using CMB, supernovae and BAO data in this way is very useful for testing and 
perhaps even ruling out some of the dark energy quintessence models.
In Section \ref{wmap} we  
consider the impact of using these new cosmological parameters on the nonlinear growth of structure.

\begin{table*}
\label{bf}
\caption{Best fit values for $\Omega_{\rm m}
h^2$, $\Omega_{\rm b}
h^2$ and $H_0$ with 68.3\% confidence intervals from minimising
$ \chi^2_{\tiny \mbox{total}}=\chi^2_{\tiny \mbox{WMAP+SN+BAO}}$
for each quintessence model. wCDM WMAP 5-year are the parameter constraints assuming a dynamical dark energy model \citep{Komatsu:2008hk}.}
\begin{tabular}{|c|r|r|r|r|}
\hline
 & $10^2\Omega_{\rm b}
h^2$ & $H_0$ (km/s/Mpc) & $\Omega_{\rm m}
h^2$& $\chi^2_{\tiny \mbox{total}}/\nu$ \\
 \hline \hline
& & & & \\
 $\Lambda$CDM WMAP 5-yr Mean & 2.267 $^{ +0.058}_{ -0.059}$ & 70.5 $\pm 1.3$ & 0.1358 $^{ +0.0037}_{ -0.0036}$ &   \\
wCDM WMAP 5-yr Mean & 2.27 $\pm 0.06$ & 69.7 $\pm 1.4 $& 0.1351 $\pm 0.0051$ &   \\
\citet{Sanchez:2009jq} & 2.267 $^{ +0.049}_{ -0.05}$ & 71.5 $\pm 1.1$ & 0.13343 $\pm 0.0026$ &  1.09 \\
\hline
& & & & \\
INV1 & 3.78 $\pm 0.145$ & 63.13 $\pm 0.5$ & 0.115 $\pm 0.0103$ & 2.27  \\
INV2 & 2.35 $\pm 0.094$ & 68.21 $\pm 0.7$ & 0.124 $\pm 0.0065$  &  1.07\\
SUGRA  & 2.68 $\pm 0.105$ &  67.63 $\pm 0.7$ & 0.111 $\pm 0.0075$ &   1.25\\
2EXP & 2.22 $\pm 0.115$ & 70.01 $\pm 0.8$& 0.138 $\pm 0.0031$ &  1.05\\
AS & 2.12 $\pm 0.121$ & 70.42 $\pm 0.9$&  0.086 $\pm 0.0121$ &  1.07\\
CNR &2.09 $\pm 0.185$ & 70.05 $\pm 1.2$ & 0.140 $\pm 0.0133$ & 1.12\\
\hline
\end{tabular}
\end{table*}

These distance priors are derived parameters which depend on the assumed cosmological model and yield constraints on dark energy parameters which are
slightly weaker than a full Markov Chain Monte Carlo (MCMC) calculation, as only part of the full WMAP data is used i.e. the $C_l$ spectrum is condensed into 2 or 3 numbers 
describing peak position and ratios and the polarisation data are ignored.
The assumed model is a standard FLRW universe with an effective number of neutrinos equal to 3.04 and  a nearly power law
primordial power spectrum with negligible primordial gravity waves and entropy fluctuations. These WMAP distance priors are extremely useful for providing
cosmological parameter constraints at a reduced computational cost compared to a full MCMC calculation.
\begin{table*}
\caption{WMAP distance priors \citep{Komatsu:2008hk} for each quintessence model using the best fit parameters $\Omega_{\rm m}
h^2$, $\Omega_{\rm b}
h^2$ and $H_0$ given in Table A3.}
\begin{tabular}{|c|r|r|r|}
\hline
 & $z*$  &  $l_A(z*)$ & $R(z*)$   \\
\hline \hline
$\Lambda$CDM WMAP 5-yr ML& 1090.51 $\pm$0.95  & 302.10 $\pm$ 0.86 & 1.710 $\pm$ 0.019   \\
Sanchez et al.2009 &  1090.12 $\pm$ 0.93   & 301.58 $\pm$ 0.67 & 1.701 $\pm$ 0.018 \\
\hline
INV1 & 1076.17 & 292.54  & 1.519   \\
INV2 &  1088.71& 301.69  & 1.676  \\
SUGRA   & 1083.96 & 298.51  & 1.596 \\
2EXP & 1091.75 & 302.91  & 1.749   \\
AS &  1087.98 & 300.23 & 1.684 \\
CNR &1093.97  & 303.51 & 1.809 \\
\hline
\end{tabular}
\end{table*}
\begin{table*}
\caption{BAO distance measurements \citep{Percival:2007yw} for each quintessence model using the best fit parameters $\Omega_{\rm m}
h^2$, $\Omega_{\rm b}
h^2$ and $H_0$
given in Table A3.}
\begin{tabular}{|c|r|r|r|r|}
\hline
& $z_{\tiny \mbox{drag}}$ & $r_s(z_{\tiny \mbox{drag}})$ & $r_s/D_V(z=0.2)$ & $r_s/D_V(z=0.35)$ \\
\hline \hline
WMAP 5-yr & 1020.5 $\pm$ 1.6 & 153.3 $\pm$ 2.0 Mpc & -&- \\
\citet{Percival:2007yw} & -& 154.758 Mpc & 0.198 $\pm$ 0.0058 & 0.1094 $\pm$ 0.0033 \\
\hline
INV1 & 1045.1 & 146.259 Mpc & 0.1765 & 0.1103 \\
INV2 & 1021.2 & 154.946 Mpc & 0.1921 & 0.1167 \\
SUGRA & 1026.4 & 155.803 Mpc & 0.1908 & 0.1161 \\
2EXP & 1019.9 & 150.983 Mpc & 0.1879 & 0.1123 \\
AS & 1010.5 & 157.745 Mpc & 0.1947 & 0.1161 \\
CNR & 1017.1 & 150.597 Mpc & 0.1876 & 0.1128 \\
\hline
\end{tabular}
\end{table*}

We shall briefly review the distance scales used in this paper and the method for finding the best fit parameters for the dark energy models.
From measurements of the peaks and troughs of the acoustic oscillations in the photon-baryon plasma
in the CMB it is possible to measure two distance ratios \citep{Komatsu:2008hk}. The first ratio
is quantified by the \lq acoustic scale\rq  , $l_A$, which is defined in terms of the sound horizon at decoupling, $r_s(z_*)$ and the
angular diameter distance to the last scattering surface, $D_A(z_*)$, as
\begin{equation}
l_A = (1+z_*) \frac{\pi D_A(z_*)}{r_s(z_*)}.
\end{equation}
Assuming a flat universe, the proper angular diameter distance is defined as
\begin{equation}
D_A(z) = \frac{c}{(1+z)}\int^z_0\frac{{\rm d}z'}{H(z')} ,
\end{equation}
and the comoving sound horizon is given by
\begin{equation}
r_s(z) = \frac{c}{\sqrt{3}}\int^{1/(1+z)}_0 \frac{{\rm d}a}{a^2H(a)\sqrt{1+(3\Omega_{\rm b}
/4\Omega_{\gamma})a}}
\label{soundhorizon}
\end{equation}
where $\Omega_{\gamma} = 2.469 \times 10^{-5}h^{-2}$ for $T_{\tiny\mbox{CMB}}= 2.725$K 
\citep{Komatsu:2008hk}
and $\Omega_{\rm b}
$ is the ratio of the baryon energy density to the critical density.
We shall use the fitting formula proposed by \citet{Hu:1995en} for the decoupling epoch $z_*$ which is a function of $\Omega_{\rm b}
h^2$ and
$\Omega_{\rm m}
h^2$ only.
The second distance ratio measured by the CMB is called the 
\lq shift parameter\rq \, \citep{Bond:1997wr}. This is the ratio of the
angular diameter distance and the Hubble horizon size at the decoupling epoch which is written as
\begin{equation}
\label{R}
R(z_*) = \frac{\sqrt{\Omega_{\rm m}
H_0^2}}{c}(1+z_*)D_A(z_*).
\end{equation}
Eq. \ref{R} assumes a standard radiation and matter dominated epoch when calculating the sound horizon.
 The expression for the shift parameter will be modified for quintessence 
models of dark energy. The proper expression for the shift parameter is given by \citep{Kowalski:2008ez}
\begin{equation}
R(z_*) = R_{\tiny \mbox{std}}(z_*)\left( \int^{\infty}_{z_*} \frac{{\rm d}z/\sqrt{\Omega_{\rm m}
(1+z)^3}}{\int_{z_*}^{\infty} {\rm d}z H_0/H(z)}\right) ,
\end{equation}
where $R_{\tiny \mbox{std}}$ is the standard shift parameter given in Eq. \ref{R}. This correction to the shift parameter can be substantial for
quintessence models with non-negligible amounts of dark energy at early times and so we include this correction for all of the scalar field models in
this paper.
The 5-year WMAP constraints on $l_A$, $R$ and the redshift at decoupling $z_*$ are the WMAP distance priors used to test models of dark energy \citep{Komatsu:2008hk}.

The angular diameter distance at the decoupling epoch can be determined from measurements of the acoustic oscillations in the
 CMB. These baryonic acoustic
oscillations are also imprinted on the distribution of matter. Using galaxies as tracers for the underlying matter distribution
the clustering perpendicular to the line of sight gives a measurement of the angular diameter distance, $D_A(z)$.
BAO data also allow us to measure the expansion rate of the universe, $H(z)$, from observations of clustering along the line of sight.
Recently, \citet{2008arXiv0807.3551G} made a 
direct measurement of the Hubble parameter as a function of redshift providing for the first time a measure of $D_A(z)$ and $H(z)$ individually. 
Using a spherically averaged correlation function
to reveal the BAO signal results in an effective distance measure given by 
 \citep{Eisenstein:2005su}
\begin{equation}
\label{DV}
D_V(z) = \left(  (1+z)^2 D_A^2(z) \frac{c z}{H(z)}\right)^{1/3}.
\end{equation}
It is the ratio of $D_V(z)$ to the sound horizon, $r_s$, at the drag epoch, $z_{\tiny \mbox{drag}}$, which determines the peak positions of the BAO signal.
The drag epoch is the redshift at which baryons are separated from photons and is slightly later than the decoupling epoch, $z_*$.
For a wide angle survey, $D_v$ is used, which is motivated on dimensional grounds and equal
sampling of all axes (e.g. $D_v$ for a pencil beam survey would have different exponents of $D_A$ and $H$).
\citet{Percival:2007yw} provide $r_s(z_d)/D_V(z)$ at two redshifts, $z=0.2$ and $z=0.35$, taken from the Sloan Digital Sky Survey (SDSS) and
Two Degree Field Galaxy Redshift Survey (2dFGRS). The two values are $r_s(z_d)/D_V(0.2)= 0.198 \pm 0.0058$ and $r_s(z_d)/D_V(0.35)= 0.1094 \pm 0.0033$.

The UNION supernovae compilation \citep{Kowalski:2008ez} consists of 307 low redshift SN  all processed using the SALT light
curve fitter \citep{2005A&A...443..781G}.
This compliation includes older data sets from the Supernova Legacy Survey and ESSENCE Survey as well as a recent dataset observed with HST.
Type Ia supernovae data is extremely useful in breaking parameter degeneracies 
such as the $w$, $\Omega_{\tiny \mbox{DE}}$ degeneracy in the CMB data.
A wide range of these two parameters can produce similar angular diameter distances at the redshift of decoupling and so SN constraints, which are almost
orthogonal to CMB constraints, help to reduce this parameter space.
The current SN data cover a wide range of redshift, $0.02 \leq z \leq 1.7$, but is only able to weakly constrain a dynamical dark energy
equation of state, $w$, at $z \geq 1$.  Also, due to a degeneracy with $\Omega_{\rm m}
$, the current SN data by themselves are not able to tightly
constrain the present value of
$w$ and including measurements involving $\Omega_{\rm m}
$ such as CMB or BAO observations break this degeneracy.

Following the prescription of \citet{Komatsu:2008hk} for using the WMAP distance priors it is necessary to find the vector $\vec{x} = (l_A, R, z_*)$
for each quintessence model in order to compute the likelihood,
$ \mathcal{ L }$, as $\chi^2 = -2\mbox{ln}\mathcal{ L } = (x_i -d_i)C^{-1}_{ij}(x_j-d_j)$,
where $\vec{d}= (l_A^{\tiny \mbox{WMAP}},
R^{\tiny \mbox{WMAP}}, z_*^{\tiny \mbox{WMAP}})$ and $C^{-1}_{ij}$ is the inverse covariance matrix for the WMAP distance priors.

In order to find the best fit cosmological parameters  for each quintessence model we minimise the function
$\chi^2_{\tiny \mbox{total}} = \chi^2_{\tiny \mbox{WMAP}} + \chi^2_{\tiny \mbox{BAO}} + \chi^2_{\tiny \mbox{SN}}$ with respect to
$\Omega_{\rm m}
h^2$, $\Omega_{\rm b}
h^2$ and $H_0$.
In appendix D of \citet{Komatsu:2008hk} it can be seen that including
the systematic errors has a very small effect on the $\Lambda$CDM parameters but can have a significant impact on dark energy parameters. Using a two
parameter equation of state for the dark energy \citet{Komatsu:2008hk} found that the parameter constraints weakened considerably after including systematic errors.
In calculating $ \chi^2_{\tiny \mbox{SN}}$ in this paper we have used
the covariance matrix for the errors on the SN distance moduli without systematic errors.

Table A1 shows the WMAP distance priors computed for each dark energy model using the cosmological parameters from \citet{Sanchez:2009jq}.
 The BAO scale
and drag redshift, $z_d$, are given in Table A2 using the same parameters. From these tables
it is clear that some
quintessence models with $\Lambda$CDM 
 cosmological parameters fail
to agree with the distance measurements within the current constraints.

With the assumption that $\Omega_{\rm m}
h^2$, $\Omega_{\rm b}
h^2$ and $H_0$ are tightly constrained by WMAP, BAO and SN data, and as a result
their posterior distribution is close to a normal distribution, minimising $\chi^2_{\tiny \mbox{total}} 
=\chi^2_{\tiny \mbox{WMAP}} +\chi^2_{\tiny \mbox{BAO}} +\chi^2_{\tiny \mbox{SN}}$ with
respect to these three parameters will be the same as marginalising the posterior distribution. We have fixed the dark energy
equation of state parameters for each quintessence model and the 68.3\% confidence intervals for each parameter
 from minimising  $\chi^2_{\tiny \mbox{total}} $
are shown in Table A3. The final column in this table is $\chi^2/\nu$ where $\nu$ is the number of degrees of freedom.
From Table A3 it is clear that the INV1 model is unable to fit the data and has a poor $\chi^2/\nu = 2.27$ statistic.
Most of the quintessence models favour a lower $\Omega_{\rm m}
h^2$ compared to $\Lambda$CDM in order to fit the distance data.
As can be seen from Table A3
the confidence intervals on the three fitted parameters $\Omega_{\rm m}
h^2$,
 $\Omega_{\rm b}
h^2$ and $H_0$ are quite large.
Once the best fit parameters from Table A3 are used, all of the quintessence models apart from 
INV1 which we rule out, 
 produce a better fit to the data, as seen in Tables A4 and A5, for the WMAP distance priors and the 
BAO distance measures respectively.
As we noted earlier the WMAP distance priors do not contain all of the WMAP power
spectrum data and only use the information
from the oscillations present at small angular scale (high multipole moments). Neglecting the Sachs-Wolfe (SW)
 effect at large angular scales (small multipole moments)
as well as polarisation data lead to weaker constraints on cosmological parameters in these dark energy models. We have  not considered
how these distance priors would change with the inclusion of dark energy perturbations \citep{Li:2008cj}.
These results are in agreement with previous work
fitting cosmological parameters of quintessence models using WMAP first year CMB data and SN data \citep{Corasaniti:2004sz}.

\bsp

\label{lastpage}

\end{document}